\documentclass[nhess, manuscript]{copernicus}
\nolinenumbers

\usepackage{bm}
\DeclareUnicodeCharacter{2060}{\nolinebreak}

\begin{document}
\title{Modeling crack arrest in snow slab avalanches -- towards estimating avalanche release sizes.}
%\title{Understanding crack arrest during dynamic crack propagation in snow slab avalanches: towards estimating the avalanche release size.}

% \Author[affil]{given_name}{surname}

\Author[1,2,3,4]{Francis}{Meloche}
\Author[3,4]{Grégoire}{Bobillier}
\Author[6]{Louis}{Guillet}
\Author[1,2]{Francis}{Gauthier}
\Author[7]{Alexandre}{Langlois}
\Author[3,4,5]{Johan}{Gaume}

\affil[1]{Laboratoire de géomorphologie et de gestion des risques en montagnes (LGGRM), Département de Biologie, Chimie et Géographie, Université du Québec à Rimouski, Canada.}
\affil[2]{Center for Nordic studies, Université Laval, Québec, Canada.}
\affil[3]{WSL Institute for Snow and Avalanche Research SLF, CH-7260 Davos Dorf, Switzerland}
\affil[4]{Climate Change, Extremes, and Natural Hazards in Alpine Regions Research Center CERC, CH-7260 Davos Dorf, Switzerland}
\affil[5]{Institute for Geotechnical Engineering, ETH Zürich, CH-8093 Zürich, Switzerland}
\affil[6]{Univ. Grenoble Alpes, Inria CNRS, Grenoble, France}
\affil[7]{Groupe de Recherche Interdisciplinaire sur les Milieux Polaires (GRIMP), Département de géomatique appliquée, Université de Sherbrooke, Canada.}

%% The [] brackets identify the author with the corresponding affiliation. 1, 2, 3, etc. should be inserted.

%% If an author is deceased, please mark the respective author name(s) with a dagger, e.g. "\Author[2,$\dag$]{Anton}{Smith}", and add a further "\affil[$\dag$]{deceased, 1 July 2019}".

%% If authors contributed equally, please mark the respective author names with an asterisk, e.g. "\Author[2,*]{Anton}{Smith}" and "\Author[3,*]{Bradley}{Miller}" and add a further affiliation: "\affil[*]{These authors contributed equally to this work.}".

\correspondence{Francis Meloche (francis.meloche@uqar.ca)}

\runningtitle{TEXT}

\runningauthor{TEXT}

\received{}
\pubdiscuss{} %% only important for two-stage journals
\revised{}
\accepted{}
\published{}

%% These dates will be inserted by Copernicus Publications during the typesetting process.

\firstpage{1}

\maketitle

%\doublespacing

\begin{abstract}
Dry-snow slab avalanches are considered to be the most difficult to predict, yet the deadliest avalanche types. The release of snow slab avalanches starts with the failure of a weak snow layer buried below cohesive snow slabs. This initial failure may propagate across the slope until the slab fractures and slides. The evaluation of crack propagation area and possible avalanche size is a primary concern for avalanche forecasters. The purpose of this study is to test the hypothesis that two primary factors may potentially stop dynamic crack propagation: the heterogeneity of snowpack properties and local terrain variations. To test this assumption, we use and further develop a depth-averaged Material Point Method (DA-MPM) which integrates MPM with shallow water assumptions for efficient elasto-plastic modeling of snow slab avalanches. Our analysis includes scenarios involving i) pure-elastic slabs and ii) elasto-plastic slabs. In the first scenario, we report a significant decrease in slab tensile stress with increasing crack speed and propose an analytical formulation explaining the stress reduction compared to quasi-static theory. In addition, we quantify the effect of weak layer heterogeneity and softening fracture energy on the crack stopping mechanism. In the second scenario, we analyse the interplay between weak layer heterogeneity and slab tensile fracture and quantify their combined effect on crack arrest. Results are interpreted through a scaling law relating the crack arrest distance to two dimensionless numbers related to weak layer strength variability and slab tensile fracture. These numbers encapsulate key parameters governing the crack arrest length, offering important insights into the size of avalanche release zones. Furthermore, the proposed model is applied to two case studies corresponding to field campaigns in which spatial variations of weak layer shear strength were measured. Based on the measurements, one can get an indication of the release size using the proposed model. Finally, DA-MPM simulations are performed on three-dimensional terrain with spatial variations either in the slope-parallel or cross-slope slope directions, revealing interesting release patterns. This research and the proposed methods can not only enhance our comprehension of the mechanical and geometrical factors influencing avalanche release sizes but also inform, in the future, the design of new or enhanced mitigation measures for avalanche start zones. 

\end{abstract}

%\copyrightstatement{TEXT} %% This section is optional and can be used for copyright transfers.

\section{Introduction}  %% \introduction[modified heading if necessary]
The release of snow slab avalanches can lead to significant volumes of fast-moving snow that can endanger people and infrastructure. It is essential to assess the size of the release area as this is a key indicator of the destructive power of the avalanche and thus a crucial parameter for risk assessment. For many years, research has been conducted to better understand the process of dry snow slab avalanches to accurately assess the stability of the snowpack \citep[e.g.][]{Fohn1987TheMechanisms,McClung1981FractureRelease,Schweizer2003SnowFormation,Monti2016SnowSnowpack}, which is necessary for avalanche forecasting. The release of a dry-snow slab avalanche requires the presence of a weak snow layer buried below cohesive snow slabs. Additional loading, such as a skier or new snowfall, can induce failure of the weak layer. The length of the failure, or crack, must exceed a so-called critical crack length to self-propagate along the slope in the weak layer. Once the propagation has started, tensile stresses build up in the slab as a result of slope-parallel displacements. Eventually, the tensile stress may reach the tensile strength of the slab, causing a slab fracture perpendicular to the weak layer, leading to the release and sliding of the slab down the slope \citep{Schweizer2016Avalanche101}. Numerous studies have been conducted on the initiation of failure and the assessment of the critical crack length needed for the onset of propagation \cite[e.g.][]{Gaume2017SnowPropagation,Gaume2018DynamicSnow,Gaume2017AssessingPropagation, Schweizer2015APrediction,Reuter2018DescribingSupport}. In the field, the assessment of the likelihood of these two processes---failure initiation and onset of crack propagation---can be done through snow mechanical tests, which can provide useful information for avalanche forecasting. However, estimating the area of the avalanche release zone remains challenging due to incomplete understanding of the mechanical processes involved. At the moment, methods based mainly on terrain characteristics and snow depth maps \citep[e.g.][]{Veitinger2016PotentialApproach,Duvillier2023DevelopmentDelineation} are used to predict potential avalanche release areas. This method yields interesting results and seems promising for operational applications, but lacks mechanical ingredients to fully account for dynamic crack propagation and crack arrest. 

Originally, some studies used a cellular automata method to focus on the influence of spatial heterogeneity of the weak layer strength on the slope stability \citep{Faillettaz2004Two-ThresholdAvalanches, Fyffe2004TheRelease, Kronholm2005IntegratingModel}. They showed that a complete failure of all cells was more likely to occur with a highly correlated spatial structure, compared to a spatial structure with random noise. These studies were the first to provide indication for the possible avalanche release size. The finite-element method (FEM) was also used to study the influence of the spatial heterogeneity of the weak layer on the avalanche release size. \cite{Gaume2015InfluenceArea} demonstrated with a FEM method that a weak layer heterogeneity could induce a slab tensile fracture that could potentially stop the propagation and release the slab. However, previously proposed cellular automata and FEM methods were unable to capture all mechanical processes involved during a dynamic crack propagation in a snow slab avalanche. Recently, snow avalanche research has shifted towards understanding dynamic crack propagation, which supposedly largely influences the avalanche size \citep{Bergfeld2023TemporalPropagation,Bobillier2021Micro-mechanicalExperiments,Bobillier2024NumericalExperiments, Trottet2022TransitionAvalanches}. Yet, so far, avalanche release size indicators accounting for the intricacies of dynamics crack propagation and arrest are still missing, urging for more research.

Crack propagation in dry snow slab avalanches was originally understood as a shear band propagation (mode II in fracture mechanics) within the weak layer \citep{McClung1981FractureRelease,Gaume2013InfluenceDepths,Puzrin2005TheSoils,Gaume2021Mechanisms1959}. However, this shear propagation model fails to reproduce crack propagation on low-angle terrain or observations of remotely triggered avalanches. \cite{Johnson2000RemotelyAvalanches} observed a change in the weak layer thickness after fracture, which indicated that the weak layer collapsed during crack propagation. He proposed a theory in which crack propagation on low-angle terrain is driven by the collapse of the weak layer and subsequent bending of the slab, which led to the introduction of a new snow mechanical test called the Propagation Saw Test (PST) \citep{Gauthier2008EvaluationLayers}. \cite{Heierli2008AnticrackAvalanches} proposed to adapt the so-called "\textit{anticrack}" theory proposed in earthquake science for crack propagation in snow avalanche. This model represents a mode of crack propagation with closing crack faces under compression (mode -I), by opposition to opening crack faces under tension (mode I). Recently, more advanced analytical models for the onset of crack propagation in avalanche release were proposed including mixed-mode weak layer failure \citep{Rosendahl2020ModelingAnticracks, Benedetti2019ATest}.

Over the past decade, various numerical simulation methods have been employed to investigate avalanche release mechanisms. Among these, the Discrete Element Method (DEM) with a bonded contact model \citep{Gaume2015ModelingMethod, Gaume2017SnowPropagation}, the finite element method (FEM) \citep{Habermann2008InfluenceRelease, Reuter2015AInstability,Gaume2018StressRelease,Podolskiy2013AMechanics} and the Material Point Method (MPM) with a new elasto-plastic law \citep{Gaume2018DynamicSnow, Gaume2019InvestigatingMethod, Trottet2022TransitionAvalanches} have been extensively utilized to simulate the initiation and dynamics of crack propagation.
Drawing from these distinct methodologies and field observations, \cite{Bobillier2021Micro-mechanicalExperiments,Bobillier2024NumericalExperiments} and \cite{Trottet2022TransitionAvalanches} documented a transition from sub-Rayleigh anticrack to supershear crack propagation during the release process. This finding was further supported by surprisingly large propagation speeds measured on avalanche videos \citep{Hamre2014FracturesAvalanches, Simenhois2023UsingSpeeds}. The study of \cite{Trottet2022TransitionAvalanches} now closes the debate on whether crack propagation is in shear mode or collapse-based anticrack mode, demonstrating that both crack propagation regimes are present in the release mechanism of a dry snow slab avalanche. On low-angle terrain or on slopes for short crack propagation distances (< 1-3 m), the propagation is driven by an anticrack mode in which the crack speed is below the shear wave speed of the slab \citep{Siron2023AAvalanches}. On steep terrain, after propagating over a distance larger than the so-called ``super critical length'', the crack transitions from a sub-Rayleigh anticrack regime to a supershear propagation regime in which the crack speed is higher than the shear wave speed of the slab. These findings suggest that a pure shear model could be enough to simulate large-scale avalanche release processes. Hence, to reduce the computational cost of 3D DEM or MPM, \cite{Guillet2023ARelease} developed a depth-averaged Material-Point-Method (DA-MPM), which simulates the supershear dynamic propagation on complex terrain in a very efficient manner.

The understanding of dynamic crack propagation is currently expanding, facilitated by enhanced measurements in the context of field PSTs \cite[e.g.][]{Bergfeld2021CrackLayers}, as well as physically-based modeling \citep{Gaume2019InvestigatingMethod,Bobillier2021Micro-mechanicalExperiments}. However, very few studies have focused on the conditions for the arrest of the crack and slab tensile failure during dynamic crack propagation \citep{Bergfeld2023TemporalPropagation, Gaume2015InfluenceArea, Gaume2019InvestigatingMethod, Trottet2022TransitionAvalanches}. More knowledge on slab tensile failure and crack arrest in a dynamic context is crucial to ultimately estimate the avalanche release size. \cite{Gaume2015InfluenceArea} made the assumption that the distance of the first slab tensile fracture could serve as an indicator for the size of the avalanche. However, their model did not explicitly represent the arrest mechanism, which could potentially be caused by a slab fracture.
\cite{Gaume2019InvestigatingMethod} demonstrated with the MPM method, two different types of slab fracture opening under tensile stress during dynamic crack propagation. The first type is related to the anticrack regime in which the bending of the slab induced an opening from the top of the slab. This type of slab fracture opening could be the cause of the crack arrests observed by \cite{Bergfeld2021CrackLayers}. They observed several crack arrests on 10 m long flat PSTs, propagating in mode -I (closing crack), at a speed lower than the shear wave speed of the slab. They suggested that these crack arrests were caused by inherent properties of the slab/weak layer system. 
In the contrary, when the crack speed is relatively high in supershear regime driven by shear, \cite{Gaume2019InvestigatingMethod} observed that the opening of the fracture started from the bottom of the slab. This bottom opening of the slab fracture was also observed in an avalanche crowns \citep{Bair2016TheResearch, Mcclung2021ApplicationRelease} suggesting shear failure of the weak layer.

Furthermore, \cite{Trottet2022TransitionAvalanches} provided insight into the effect of slab tensile fractures on crack propagation speed. They reported that slab fractures could prevent propagation from reaching the supershear speed if they occur before or during the transition between the anticrack and supershear regime. Multiple slab fractures could potentially slow the crack and cause a crack arrest. They also showed that once the crack is in the supershear regime, i.e. at a crack speed higher than the shear wave speed of the slab, the slab fractures do not affect the crack speed any longer. This result could potentially indicate that supershear crack propagation could only be stopped if something changes in the system, either the topography or a heterogeneity in the snow properties. Further work is needed to test this hypothesis.

In accordance with the arguments put forth by \cite{Guillet2023ARelease} which draw from the findings of \cite{Trottet2022TransitionAvalanches} regarding the supershear transition in avalanche release, and considering the significant computational expenses associated with 3D simulation tools like MPM or DEM, we have chosen to investigate crack arrest mechanisms during dynamic crack propagation using the depth-averaged Material Point Method (DA-MPM). This approach was recently introduced and verified by Guillet et al. (2023). DA-MPM enables rapid calculations over large areas, allowing to simulate these avalanche releases on realistic slopes within few minutes instead of days. This enables an investigation on potential drivers of crack arrest at the slope scale. Our work focuses on crack arrest that occurs when dynamic crack propagation reaches a steady state in the supershear regime, at a crack speed near the longitudinal elastic wave speed of the slab. Our work is based on the following hypothesis, which is based on the results of \cite{Trottet2022TransitionAvalanches} and \cite{Simenhois2014ObservationsBoundaries}: we assume only two drivers can stop a crack propagation in a dynamic supershear mode: 1) an internal heterogeneity in the snowpack, 2) an external variation such as topography.

In this paper, we aim to quantify the effects of weak layer heterogeneity and slab fracture on crack arrest propensity. These effects are examined both independently and in combination. In Section 2, we introduce the numerical method employed, namely the Depth-Averaged Material Point Method (DA-MPM), along with the simulation setup. In subsequent sections, we present results for two scenarios: first, for purely elastic slabs, and second, for elasto-plastic slabs allowing for slab tensile fractures. We conduct a sensitivity analysis and propose a scaling law to elucidate crack arrest distances based on two dimensionless numbers. Furthermore, we present an application based on real weak layer heterogeneity derived from field measurements and outline future directions, including the extension of our study to 3D simulations at the slope scale. Finally, in Section 5, we discuss the results and their implications, highlighting the model limitations and avenues for further research.

\section{Methods}
\subsection{DA-MPM}
\begin{figure}
\includegraphics[width=0.8\textwidth, trim = 1cm 10cm 8cm 0cm]{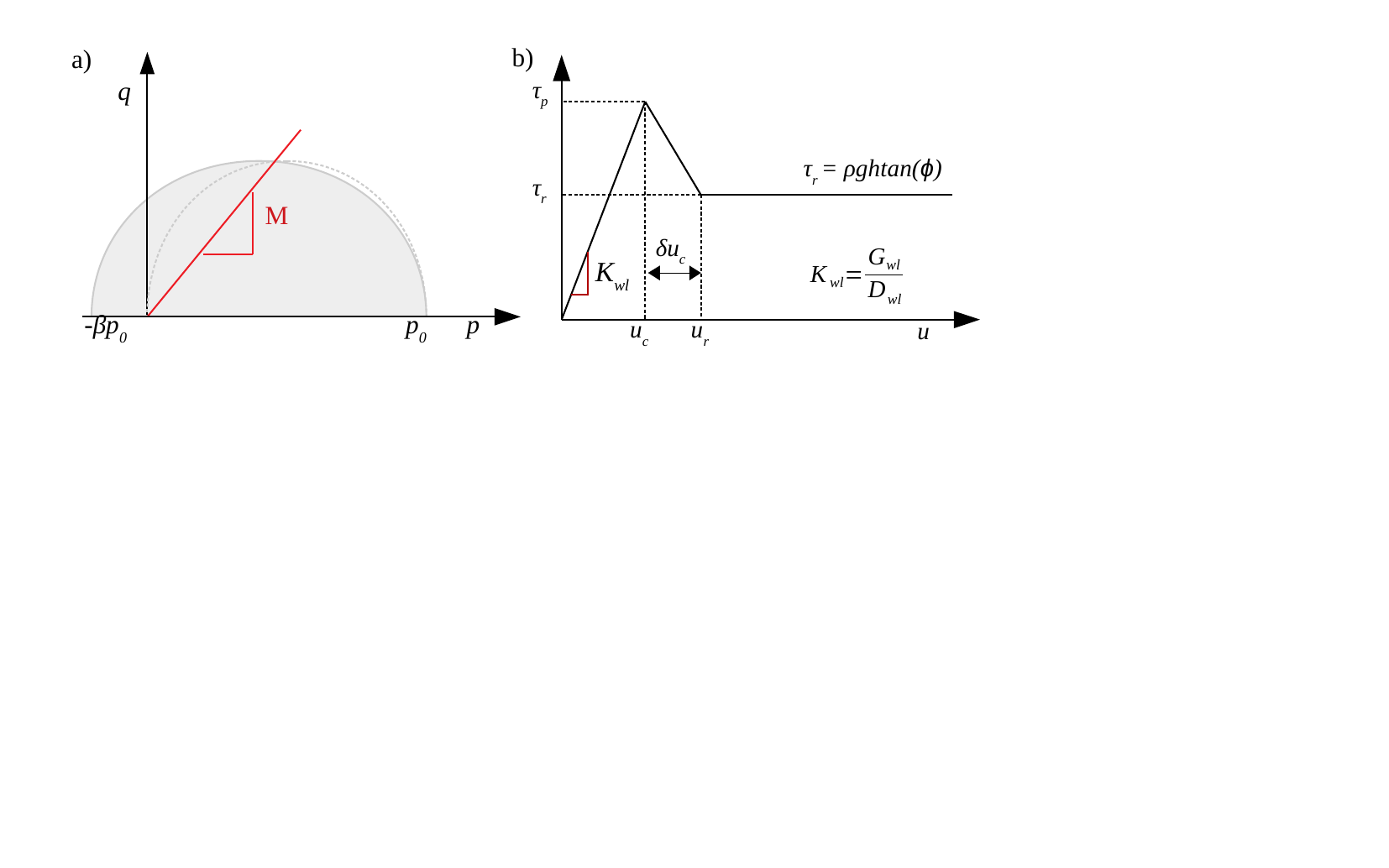}
\caption{a) Slab elasto-plastic model following a Modified Cam-Clay yield surface, with $q$ representing the Von Mises equivalent stress, $p$ the pressure, $M$
the slope critical state line, $p_0$ the compressive strength, and the $\beta p_0$ the tensile strength. b) The weak-layer is defined as as quasi-brittle interface model with the peak shear stress $\tau_p$ at the critical displacement $u_c$, and the residual stress $\tau_r$ at the residual displacement $u_r$. The residual stress is defined as $\tau_r=\rho g h \tan \phi$ where $\rho$ is the particle density, $g$ is the gravitational acceleration, $h$ the particle height and $\phi$ is the friction angle. The softening behavior is defined by the softening coefficient $\delta$.}
\label{fig:MCC}
\end{figure}

The findings of \cite{Trottet2022TransitionAvalanches} support our choice to adopt a depth-averaged approach that solves the momentum and mass conservation equations. The main assumptions will be outlined in this section, however, for a more detailed presentation of the method, please refer to \cite{Guillet2023ARelease}.
The DA-MPM method is based on the classic shallow water assumption, i.e the flow thickness is much smaller than slope-parallel dimensions of the flow. It can be translated into our avalanche context as ${h}/{L} \ll 1 $, where $h$ is the slab height and $L$ is the characteristic length of the avalanche. The material in this study is assumed to be incompressible, meaning that the density $\rho$ is constant regardless of time, place, or position. The flow surface is stress free at the top of the slab, where $\boldsymbol{\sigma_}{h=z} = \boldsymbol{0}$. It should be noted that the method can be easily adapted to a compressible form if the evolution of $\rho$ is necessary. In contrast to classical MPM in which the continuum is discretized using particles, in DA-MPM, colums of height $h$ are utilized  (Fig.\ref{fig:MCC}). The depth-averaged equation of mass conservation is given by:
\begin{equation}
    \frac{\partial h}{\partial t} + \frac{\partial (h \overline{v}_x)}{\partial x} + \frac{\partial(h\overline{v}_y)}{\partial y} = 0
\end{equation}
where $\overline{v}_x$ and $\overline{v}_y$ are the depth-averaged velocities at the integration point of the particle.

The last assumption is a plane stress assumption with $\sigma_{zy} = \sigma_{zx} = 0$. Basal forces are applied at the interface between the slab and the weak layer defined by:
\begin{equation}
    \tau_{xz} : = \sigma_{xz}|_{z=0}, \tau_{yz}:= \sigma_{yz}|_{z=0}
\end{equation}
which give also $\sigma_{zz}|_{z=0} = \rho g_z h$ at the bottom of the slab column and $\overline{\sigma}_{zz} = \frac{1}{2}h\rho g_z $ at the integration point, located in the middle of the column. This allows us to have the depth-averaged non-conservative form of momentum conservation:
\begin{equation}
    \rho h \frac{\overline{d}\overline{v}_x}{\overline{d}t} = \frac{\partial(h \overline{\sigma}_xx)}{\partial x} + \frac{\partial(h \overline{\sigma}_xy)}{\partial y} - \tau_{xz} + g_x \rho h .
\end{equation}

\begin{equation}
    \rho h \frac{\overline{d}\overline{v}_y}{\overline{d}t} = \frac{\partial(h \overline{\sigma}_yy)}{\partial y} + \frac{\partial(h \overline{\sigma}_xy)}{\partial x} - \tau_{yz} + g_y \rho h .
\end{equation}
where the depth-averaged material derivative ${\overline{d}}/{\overline{d}t}$ is given by ${\overline{d}}/{\overline{d}t}={{\partial}}/{{\partial}t}+\overline{v}_x{{\partial}}/{{\partial}x}+\overline{v}_y{{\partial}}/{{\partial}y}$.

The slab is represented with an elasto-plastic model following a Modified Cam-Clay yield surface $\gamma$ in the $p-q$ space \citep{Gaume2019InvestigatingMethod}:
\begin{equation}
    \gamma(p,q) = q^2(1+2\beta) + M^2(p + \beta p_0)(p-p_0)
\end{equation}
where $M$ is the cohesionless critical state line, $\beta$ is the cohesion parameter that quantifies the ratio between the tensile and compressive resistance and $p_0$ is the consolidation pressure that affects the size of the yield surface. With $M$ and $\beta$ constant, respectively, at 0.3 and 1.2 \citep{Guillet2023ARelease}, we related the uniaxial slab tensile strength $\sigma_t$ to $p_0$ according to:
\begin{equation}
    p_0 = \frac{-(1-\beta)p_t}{2 \beta}  + \sqrt{\left(\frac{(\beta + 1)p_t}{2\beta}\right)^2 + \left(\frac{q_t}{M}\right)^2 \left(\frac{(1+2\beta}{\beta}\right)}
\end{equation}
where $p_t = \frac{\sqrt{3}\sigma_t}{2}$ and $q_t = -\frac{\sigma_t}{2}$ considering the plane stress hypothesis (see Appendix B). The weak layer is conceptualized as a quasi-brittle interface that exhibits a softening behavior when the displacement $u$ reaches the critical values $u_c$, corresponding to the shear stress reaches $\tau_p$ (Figure \ref{fig:MCC}-b). The stress $\tau$ increases with the displacement $u$ until $\tau_p$. The associated stiffness $K_{wl}$ can be related to the shear modulus of the weak layer $G_{wl}$ and weak layer thickness $D_{wl}$ according to $K_{wl}=G_{wl}/D_{wl}$. The softening continues until the displacement reaches its residual values $u_r$ , after which the stress levels-off at a residual frictional value $\tau_r$. The residual friction is given by the slab density $\rho$, the slab thickness $h$, the gravity acceleration $g$ and friction angle of snow $\phi = 27^\circ$. The residual displacement $u_r$ is given by $u_r = u_c + \delta u_c$. The softening behavior is defined by the softening coefficient $\delta$, so that a high $\delta$ will have a important softening behavior in the weak layer (a value of $\delta=0$ corresponds to a perfectly brittle behavior). 

%% [trim={left bottom right top},clip]
\begin{figure}
\includegraphics[width=0.4\textwidth, trim = 8cm 9cm 12cm 0cm]{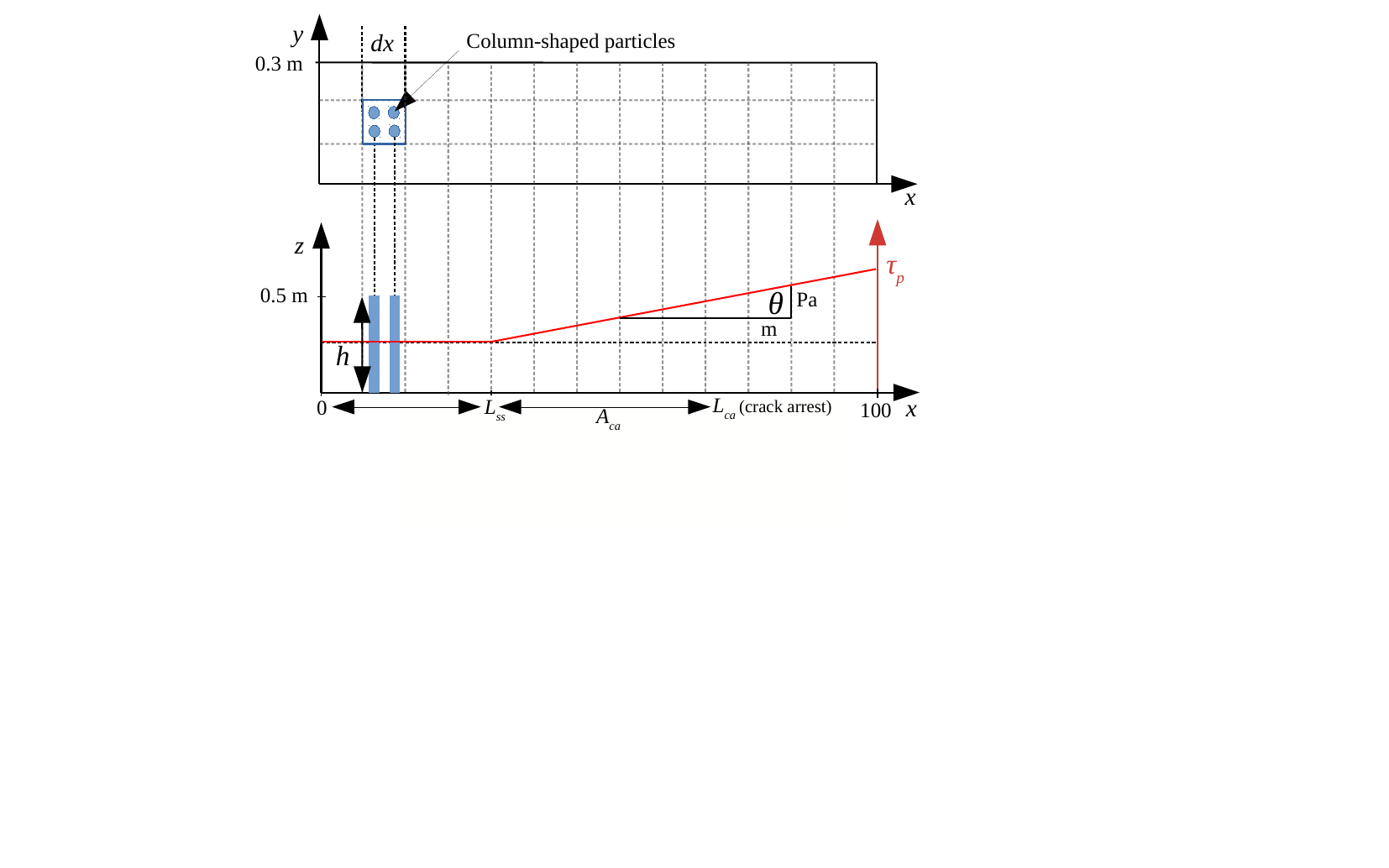}
\caption{Geometry of the PST’s simulation in ($x$, $y$) (a, top view) and ($x$, $z$) (b, side view) planes. The particle height $h$ is constant at 0.5 m. The simulations with the weak layer strength variation (red) were defined with the shear strength gradient $\theta$ is defining the shear strength $\tau_p$ beyond $L_{ss}$ of the PST. The crack arrest position $L_{ca}$ is where the crack speed has returned to zero. The crack arrest length $A_{ca}$ is defined as $L_{ca}-L_{ss}$.}
\label{fig:PSTgeometry}
\end{figure}

We used DA-MPM to simulate PSTs which are 100 m long and 30 cm wide, with a $dx$ set to 0.02 m on with a slope angle $\psi=35^\circ$. The initial column height $h$ (slab thickness) was set to 0.5 m and the slab density was set to 250 kg m$^{-3}$. As the slab elastic modulus $E$ crucially impacts crack propagation dynamics, it will be varied within a realistic range and its effect on the results will be analyzed in details. The goal of this study is to test our research hypothesis that a internal snow variation could lead to a crack arrest in supershear regime. The internal snow variation was represented by the shear strength of the weak layer $\tau_p$, with a constant and positive shear strength gradient $\theta$. A propagation towards stronger zones of the weak layer (increase of shear strength) appears appropriate in a real avalanche context because the fracture initiation induced by a trigger will typically start within weaker zones of the snowpack. For these simulations, all snow properties ($h$, $\rho$, $E$, $\sigma_t$) were constant across the PST except for $\tau_p$, which was constant at 1000 Pa over a certain distance called the steady-state length $L_{ss}$ and then gradually increases according to a shear strength gradient $\theta$ in Pa m$^{-1}$ (Figure \ref{fig:PSTgeometry}). The steady-state length $L_{ss}$ is the distance that ensures the crack propagation speed reaches a steady-state regime. To simulate a PST, we numerically removed the cohesion of the weak layer, and applied the residual friction $\tau_r$ (Figure \ref{fig:MCC}-b). The removal of the weak layer at the bottom of the PST will induce stress concentration at the crack tip, until it reached the stress necessary for the onset of crack propagation. Then, the crack dynamically propagates as a shear band up-slope until a crack arrest was recorded or until the crack propagated across the entire PST without arrest. The crack speed was normalized by the the shear wave speed of the slab given by $c_s = \sqrt{\frac{G}{\rho}}$, with $G$ being the slab shear modulus and $\rho$ the slab density. The slab shear modulus is defined by $G = \frac{E}{2(1+\nu)}$, with $\nu$ being the Poisson's ratio set to 0.3 in all simulations.

Four different types of simulations were performed in this study that will be presented in distinct result sections. 
First, simulations will be performed to investigate an the effect of crack speed on slab tensile stress. For these simulations, different controlled crack speed are used to investigate the dynamic tensile stress build up in the slab, and compared to the quasi-static tensile stress, given by $\sigma^{qs}_{xx} = \rho g a \sin(\psi)(1-\frac{\tan(\phi)}{\tan(\psi)})$, where $a$ is the crack length growing during the simulation. These PSTs were conducted with a pure elastic slab and an infinite shear strength, making self-sustained dynamic crack propagation not possible. Hence, a "controlled" crack propagation was enforced by numerically removing the cohesion of the weak layer at different and constant speed values. This controlled procedure enabled us to illustrate the effect of crack speed on dynamic tensile stress in the slab.

\begin{table*}[t]
\caption{Summary of all parameters for the four different types of simulations performed.}
\label{table.param}
\begin{tabular}{c c c c c}
\tophline
Parameters & PST "controlled" speed & Pure elastic slab  & Elasto-plastic slab & Full slope scale \\
\middlehline
Column/slab height $h$ & 0.5 m & 0.5 m & 0.5 m & 0.5 m \\
Column/slab density $\rho$ & 250 kg m$^{-3}$ & 250 kg m$^{-3}$ & 250 kg m$^{-3}$ & 250 kg m$^{-3}$ \\
Weak layer shear strength $\tau_0$ & $\infty$ Pa & 1000 Pa & 1000 Pa & mean 1000 Pa \\
Weak layer shear modulus $G_{wl}$  & 0.2 MPa & 0.1...0.3 MPa & 0.2 MPa & 0.2 MPa \\
Weak layer thickness $D_{wl}$ & 0.04 m & 0.04 m & 0.04 m & 0.04 m \\
Poisson's ratio $\nu$ & 0.3 & 0.3 & 0.3 & 0.3 \\
Slope angle $\psi$ & 35° & 35° & 35° & max 35° \\
Internal snow friction angle $\phi$ & 27° & 27° & 27° & 27° \\
PST sawing speed  & 0.01...1.6 $c_s$ & 6 m s$^{-1}$ = 0.07 $c_s$ & 6 m s$^{-1}$ = 0.07 $c_s$ & 6 m s$^{-1}$ = 0.07 $c_s$ \\
\middlehline
Steady-state length $L_{ss}$ & NA & 5...30 m & 10-20-30 m & NA \\
Softening coefficient $\delta$ & 0 & 0...4  & 0-1 & 1 \\
Shear strength gradient $\theta$ & NA & 100...10 000 Pa m$^{-1}$  & 10...1000 Pa m$^{-1}$ & 20 Pa m$^{-1}$* \\
Tensile strength $\sigma_t$ & $\infty$ kPa & $\infty$ kPa & 2..10 kPa & 4 kPa \\
Slab elastic modulus $E$ & 4 MPa & 2-4-6 MPa  & 2-4-6 MPa & 4 MPa \\
\bottomhline
\end{tabular}
\belowtable{*GRF Gaussian Random field was used to generate 2D surface gradient of shear strength gradient $\theta$.} % Table Footnotes
\end{table*}

Second, a sensitivity analysis was conducted with a pure elastic slab. A pure-elastic slab, although non-realistic, will allow us to test the following parameters in a simpler case study to isolate important mechanical drivers: the steady-state length $L_{ss}$, the softening coefficient $\delta$, the shear strength gradient $\theta$, the weak layer shear modulus $G_{wl}$, and the slab elastic modulus $E$. 

Third, the PST analysis was made with an elasto-plastic slab. This allowed us to study the effect of slab fracture(s) on the propagation speed and possibly the arrest of the crack. This analysis involves one additional parameter: the slab tensile strength $\sigma_t$. In addition, we performed similar simulations with \textit{in-situ} shear strength values and gradients obtained based on field measurements conducted by \cite{Meloche2024SnowModelling}. Using a high-resolution penetrometer named SnowMicroPen \citep{Johnson1999CharacterizingSnow}, they mapped the shear strength of the weak layer across various avalanche-prone slopes.

The last type of simulation presented in this research represents an application to predict the avalanche release size at the slope scale over three-dimensional terrain with a spatial variability of the weak layer strength. The slope scale simulation was performed on two generic slopes with dimensions 50 m (down-slope) by 60 m (cross-slope) for the first slope and 50 m (down-slope) by 100 m (cross-slope) for the second slope, aimed at studying cross-slope propagation. The generic slope consists of a concave zone at the slope toe which reaches a maximum slope angle $\psi$ = 35°, followed by a convex shape at the top of the slope. A Gaussian random field GRF approach was used to generate the shear strength heterogeneity with similar strength gradients as those measured experimentally. Table \ref{table.param} lists all parameter values used for the four different types of simulation, including the range values tested of each parameters. Finally, we define $A_{ca}$, the crack arrest length which is computed as $A_{ca}=L_{ca}-L_{ss}$, where is $L_{ca}$ is the $x$-position of the crack arrest in the coordinate system defined in Figure \ref{fig:PSTgeometry}.

\begin{figure}
\includegraphics[width=0.9\textwidth]{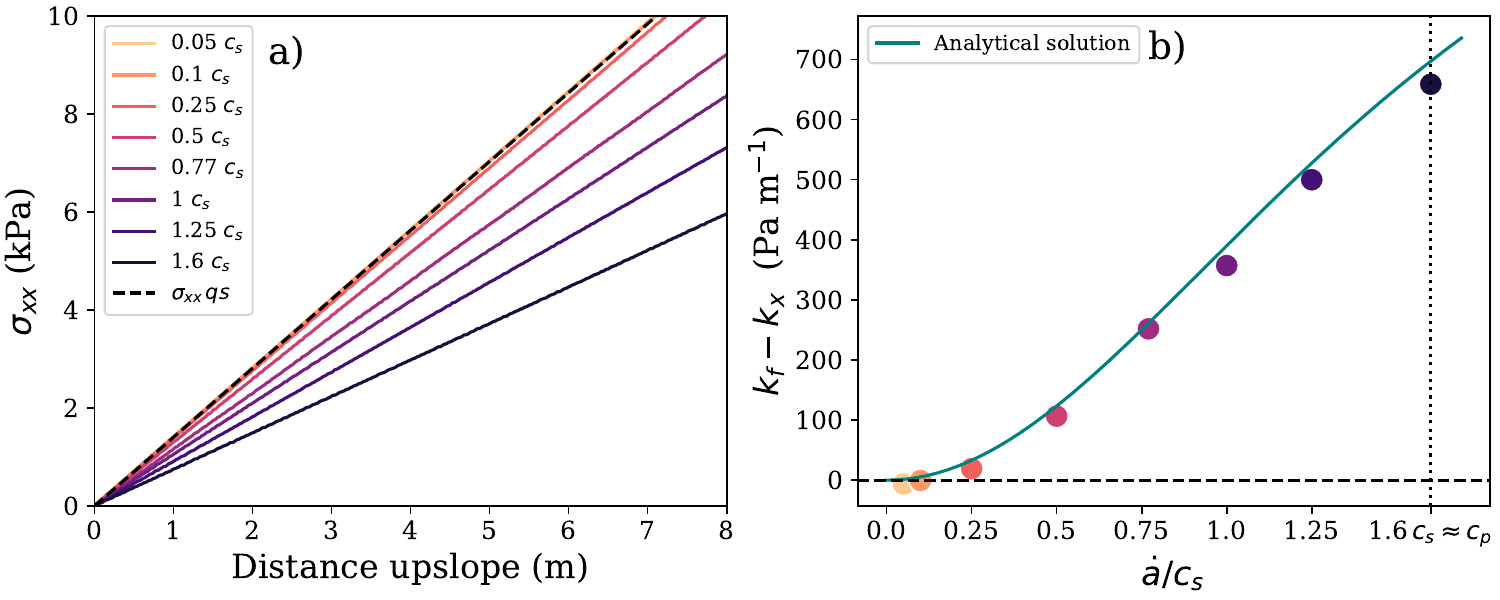}
\caption{a) Tensile stress $\sigma_{xx}$ as a function of the distance upslope in a PST simulation with infinite weak layer shear strength in a controlled sawing procedure. The numerical sawing is performed by removing weak layer cohesion at different speeds, along the upslope direction. The theoretical quasi-static tensile stress $\sigma^{qs}_{xx}$ is also plotted for reference. b) Difference between the quasi-static tensile stress gradient $k_f$ (Eq. (7)), and the dynamic tensile stress gradient $k_x$ (Eq. (8)) versus the normalized sawing speed $\dot a/c_s$, until $c_p \approx 1.6 c_s$.}
\label{fig:undershoot}
\end{figure}

\section{Results}
\subsection{Dynamic effects on slab tensile stress}
An in-depth analysis of our simulations revealed that the tensile stress was significantly lower than the quasi-static tensile stress given by $\sigma^{qs}_{xx} = a\rho g \sin(\psi)\left(1-\frac{\tan(\phi)}{\tan(\psi)}\right)$, where $a$ is crack length increasing during the simulation. We investigated this reduction in tensile stress induced by dynamic effects using a PST setup with ''controlled'' speed (i.e. the crack is artificially created with a constant speed). This controlled procedure enabled us to quantify the effect crack speed on slab tensile stress. Figure \ref{fig:undershoot}-a shows the tensile stress as a function of the distance upslope along the PST. The black dotted line represents the quasi-static tensile stress. As the speed increases, the tensile stress deviates from the quasi-static formulation.
Figure \ref{fig:undershoot}-b demonstrates the difference between the quasi-static tensile stress gradient $k_f$ and the dynamic tensile stress gradient $k_x$ for different normalized speed values $\dot a/c_s$. The increase of the tensile stress gradient difference $k_f-k_x$ with crack speed is well reproduced by a newly developed analytical expression (Figure \ref{fig:undershoot}-b) given by Eqs. (7) and (8): 
\begin{equation}
    k_f = \rho g \sin(\psi)\left(1-\frac{\tan(\phi)}{\tan(\psi)}\right)
\end{equation}
and,
\begin{equation}
    k_x = \frac{c_p^2 k_f}{c_p^2 + \dot a^2} 
\end{equation}
where $c_p = \sqrt{\frac{E^\prime}{\rho}}$ represents the elastic wave speed of the slab, $E^\prime = E/(1-\nu^2)$, and $\nu = 0.3$ is the Poisson's ratio. The derivation of the analytical solution is detailed in Appendix A. 

\begin{figure}
\includegraphics[width=0.8\textwidth]{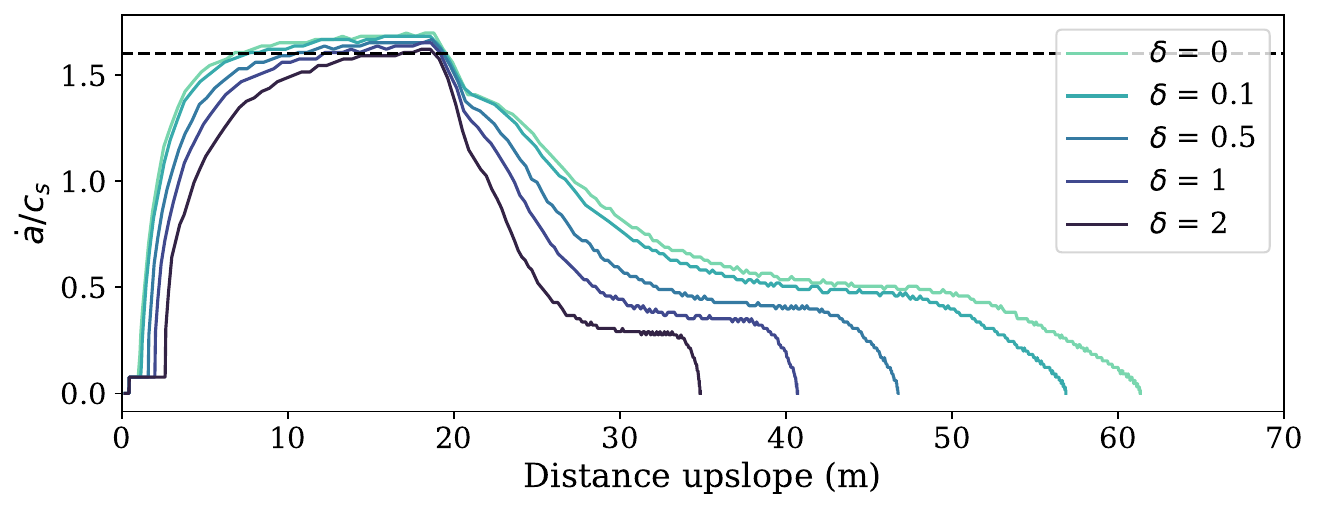}
\caption{Crack speed $\dot a$ along the PST, normalised by the slab shear wave speed $c_s = \sqrt{G/\rho}$. The colors represent different simulations with specific softening coefficients $\delta$ (0 = brittle). Crack arrest is observed when the crack speed goes back to zero. A horizontal dashed line is set at 1.6 $c_s \approx \sqrt{{E^\prime}/{\rho}}$, which is the supershear speed \citep{Trottet2022TransitionAvalanches}.}
\label{fig:PSTdelta}
\end{figure}

\subsection{Simulations with Elastic Slabs}
The first simulations were performed with a pure elastic slab to isolate important drivers with potentially five parameters that could affect the crack arrest length. First, we analyze the effect of the softening coefficient $\delta$ which relates to the energy dissipated during fracture. Figure \ref{fig:PSTdelta} shows five PST simulations with different softening coefficients $\delta$ with the same shear strength gradient. First, the crack speed reached 1.6 $c_s$, within the steady-state length ($L_{ss}$ = 20 m). Different PST simulations exhibited the same behavior within $L_{ss}$, reaching supershear speed at 1.6 $c_s$, except that the simulations with the lower $\delta$ values (pure brittle) reached 1.6$c_s$ earlier than the simulations with higher $\delta$ (Figure \ref{fig:PSTdelta}). Beyond $L_{ss}$, the crack speed was significantly influenced by the shear strength gradient ($\theta$ = 500 Pa m$^{-1}$), eventually causing the crack speed to return to zero. All PST simulations displayed a rapid deceleration in the crack speed, followed by a period of rather constant crack speed, before returning to zero (Figure \ref{fig:PSTdelta}). This sustained constant crack speed observed at the end of crack propagation was a result of the combined effects of deceleration at the crack tip and the pure elastic slab, generating residual downward stress leading to a "late push" in the crack propagation. As will be seen later, this finding may not applicable to rather soft elasto-plastic snow slabs for which a slab fracture would have occurred before. Finally, the effect of the softening coefficient was demonstrated with the pure brittle simulation ($\delta$ = 0) with the longest crack arrest length $A_{ca}$ of 62 m, compared to a $\delta$ of two with almost half the length at 35 m (Figure \ref{fig:PSTdelta}). This finding demonstrated how increasing weak layer fracture energy resulted in shorter $A_{ca}$. 

\begin{figure}
\includegraphics[width=0.9\textwidth]{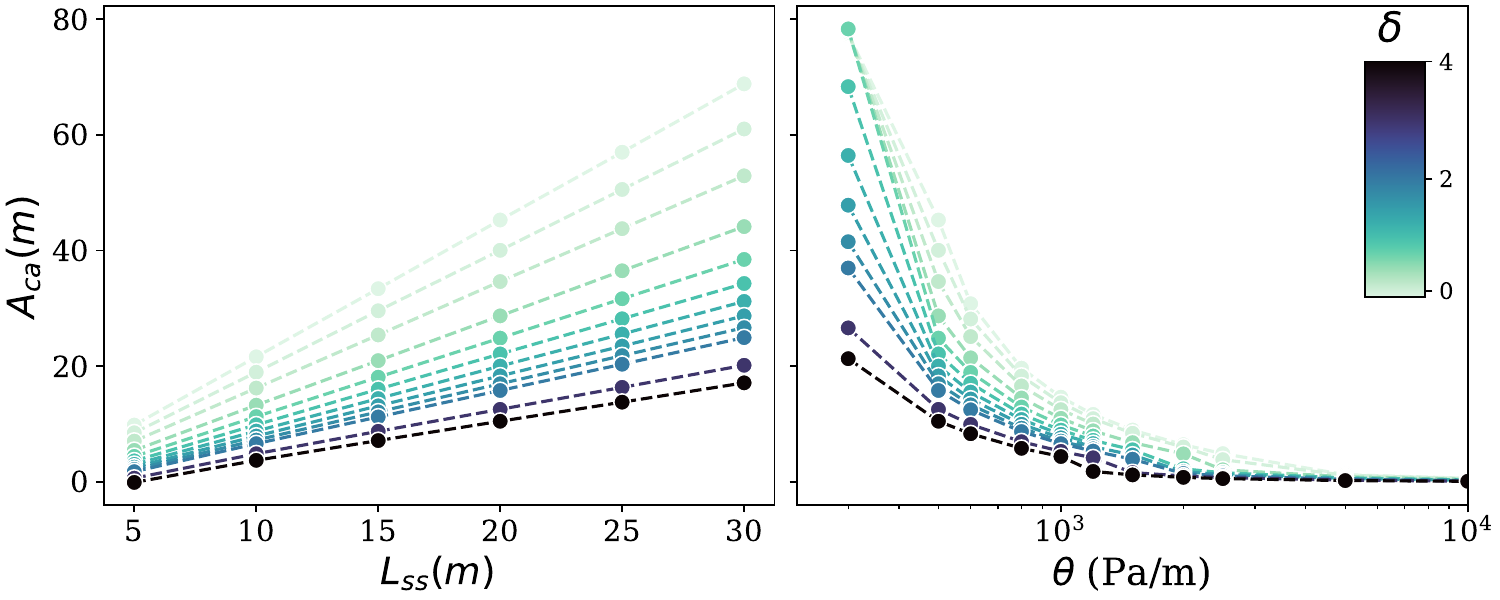}
\caption{Crack arrest length $A_{ca}$ from the PST simulations in relation to a) the super shear length $L_{ss}$ with different softening coefficients $\delta$, and a constant $\theta = 500$ Pa m$^{-1}$ and to b) the shear strength gradient $\theta$ with different softening coefficient $\delta$, and a constant $L_{ss} = 20 m$.}
\label{fig:PSTelastic_thetaLss}
\end{figure}

The steady-state length $L_{ss}$ is the second parameter that could potentially affect the crack arrest length. Several PST simulations with different values of $L_{ss}$ demonstrated a linear effect of $L_{ss}$ over $A_{ca}$ (Figure \ref{fig:PSTelastic_thetaLss}). Indeed, slabs with cracks which propagated over longer distances $L_{ss}$ at a speed around 1.6 $c_s$ stored more elastic energy than those with lower $L_{ss}$ values. Higher stored elastic energy also implies higher energy to dissipate to stop crack propagation thus naturally leading to a higher $A_{ca}$ value. Another parameter that obviously significantly affects the crack arrest length is the shear strength gradient $\theta$. His effect on $A_{ca}$ is nonlinear as $A_{ca}$ rapidly decreases as $\theta$ increased, especially around 500 Pa m$^{-1}$ (Figure \ref{fig:PSTelastic_thetaLss}). When the shear strength gradient is around 5000 Pa m$^{-1}$, the gradient is too large, causing crack arrest after $L_{ss}$. Figure \ref{fig:PSTelastic_thetaLss} also illustrates the influence of the softening coefficient $\delta$ on the crack arrest length, in combination with the effects of $L_{ss}$ and $\theta$.

\begin{figure}
\includegraphics[width=0.9\textwidth]{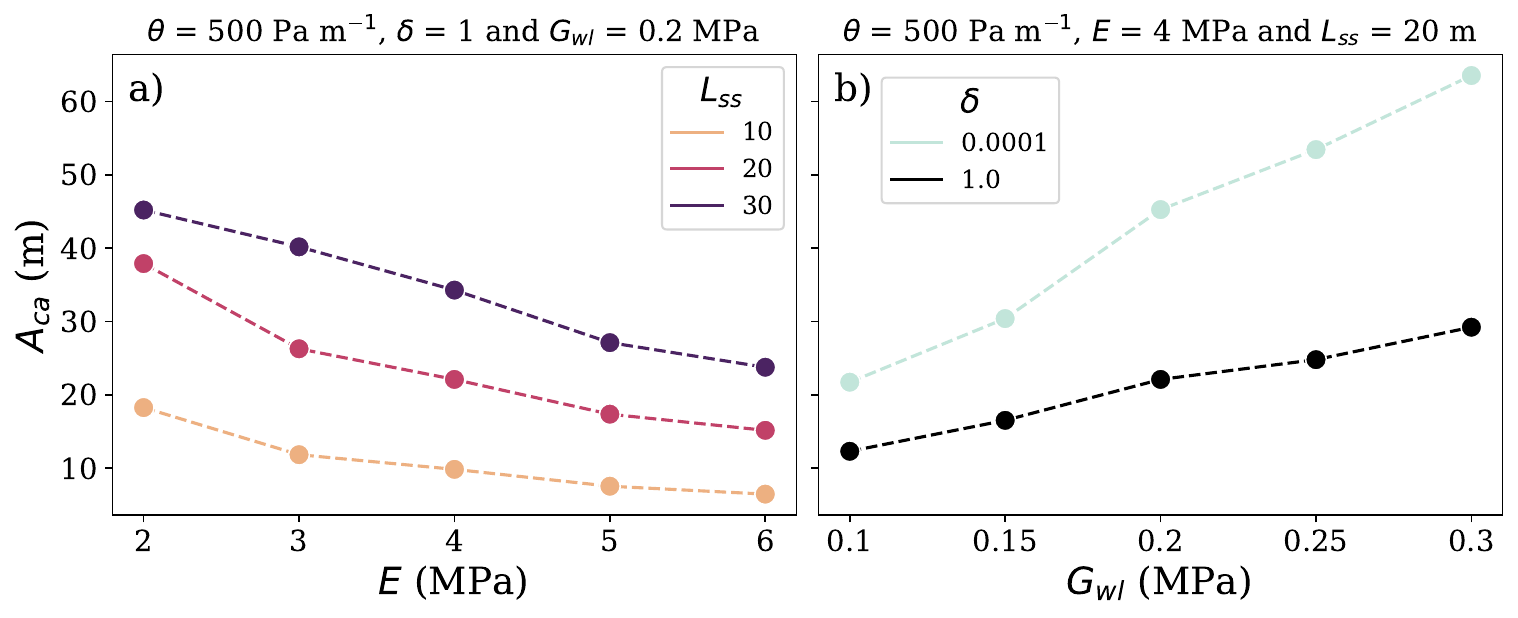}
\caption{Crack arrest length $A_{ca}$ from the PST simulations in relation to a) different elastic modulus of the slab $E$ with different steady-state length $L_{ss}$, a constant $\theta = 500$ Pa m$^{-1}$, softening coefficient $\delta$ = 1 and a shear modulus $G_{wl}$ = 0.2 MPa and in relation to b) the weak layer shear modulus $G_{wl}$, with different softening coefficient $\delta$, and a constant $L_{ss}$ = 20 m, shear strength gradient $\theta = 500$ Pa m$^{-1}$, and slab elastic modulus $E$ = 4 MPa.}
\label{fig:PSTelastic_EGwl}
\end{figure}

The last two parameters that were tested with a pure elastic slab are the slab elastic modulus $E$ and the weak layer shear modulus $G_{wl}$. These two parameters have a significant, yet opposite impact on the crack arrest length $A_{ca}$. Indeed, both parameters affect the characteristic elastic length of the system $\Lambda= \sqrt{\frac{E h D_{wl}}{(1-\nu^2)G_{wl}}}$. If the slab elastic modulus of the slab increases, $\Lambda$ also increases, thus reducing stress concentrations at the crack tip, resulting in a reduction in crack arrest length (Figure \ref{fig:PSTelastic_EGwl}-a). On the other hand, if the the shear modulus of the weak layer increases, $\Lambda$ decreases resulting in higher stress concentrations and thus a higher value of $A_{ca}$.

\begin{figure}[h!]
\includegraphics[width=0.85\textwidth]{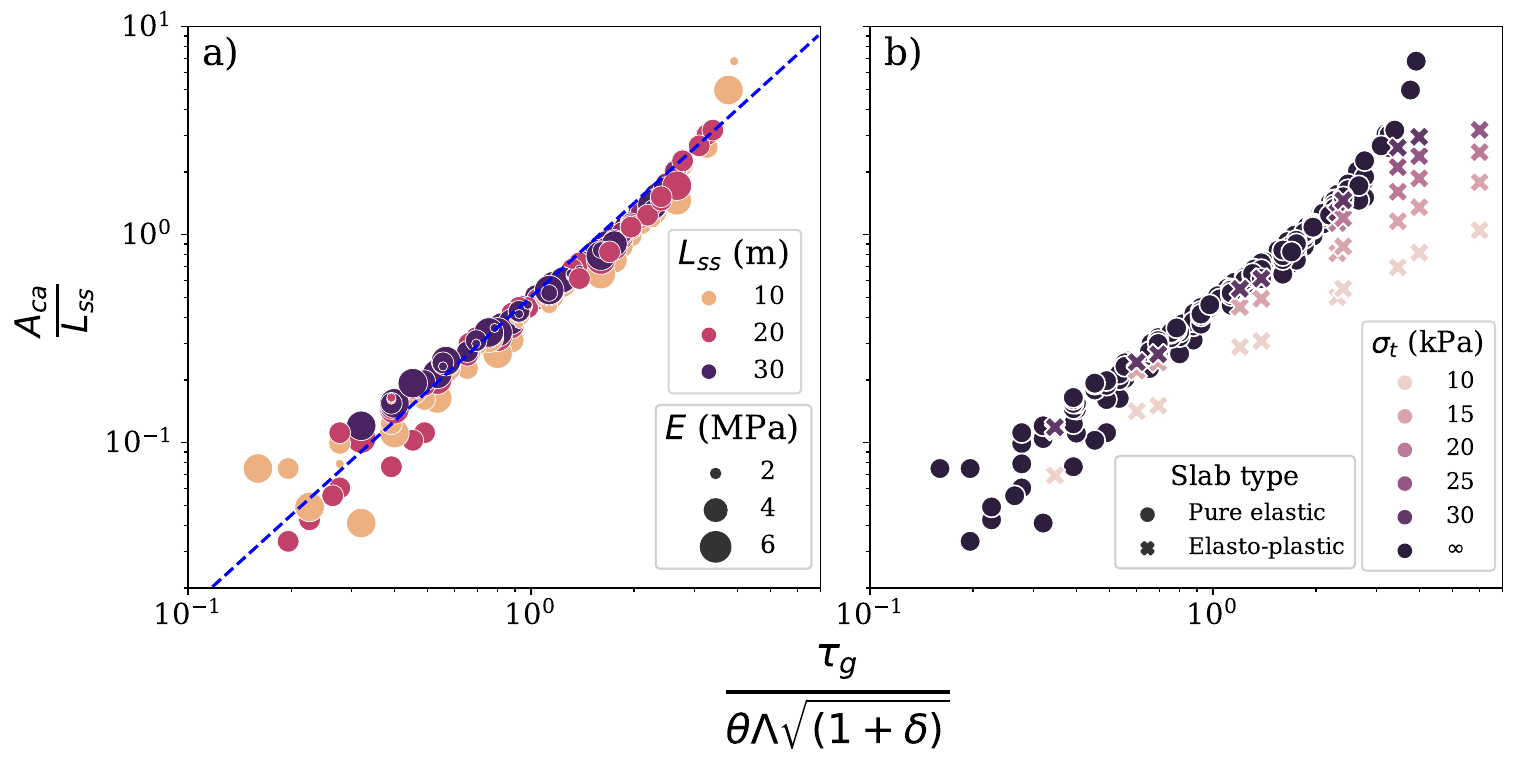}
\caption{Crack arrest length $A_{ca}$ (normalized with $L_{ss}$) from all our PST pure elastic slab simulations in relation to the dimensionless number $\tau_g/(\theta\Lambda\sqrt{1+\delta})$. The PST simulations presented in a) were all pure elastic slab simulations.  The 2:3 line is plotted in blue as reference. In b) the same PST simulations as in a) are shown in dark blue but with additional PST elasto-platic simulations with different values of $\sigma_t$. The data in this plot includes the results from all pure elastic simulations with the following parameters that were varied: the shear strength gradient $\theta$, the weak layer shear modulus $G_{wl}$, the steady-state length $L_{ss}$, elastic modulus of the slab $E$, and the softening coefficient $\delta$.}
\label{fig:PSTaca_purescale}
\end{figure}

Following a comprehensive examination of the simulation data, a scaling law relating the dimensionless crack arrest length $A_{ca}/L_{ss}$ to the dimensionless variable $\tau_g/(\theta\Lambda\sqrt{1+\delta})$ was found (Figure \ref{fig:PSTaca_purescale}a). This number basically compares a driving contribution, namely the shear stress $\tau_g$ to a resisting term, namely the shear strength increment $\theta\Lambda$ over a distance $\Lambda$ (characterizing elastic stress concentrations) multiplied by $\sqrt{1+\delta}$ (characterizing the softening distance in the weak layer mechanical model). On the basis of this data collapse and linear character of the relationship in log-scale, we can provide the following empirically-based power-law equation:
\begin{equation}
    \frac{A_{ca}}{L_{ss}}\propto \left(\frac{\tau_g}{\theta\Lambda\sqrt{1+\delta}}\right)^{3/2}
\end{equation}

\subsection{Simulations with Elasto-Plastic slabs}
\begin{figure}[h!]
\includegraphics[width=0.83\textwidth]{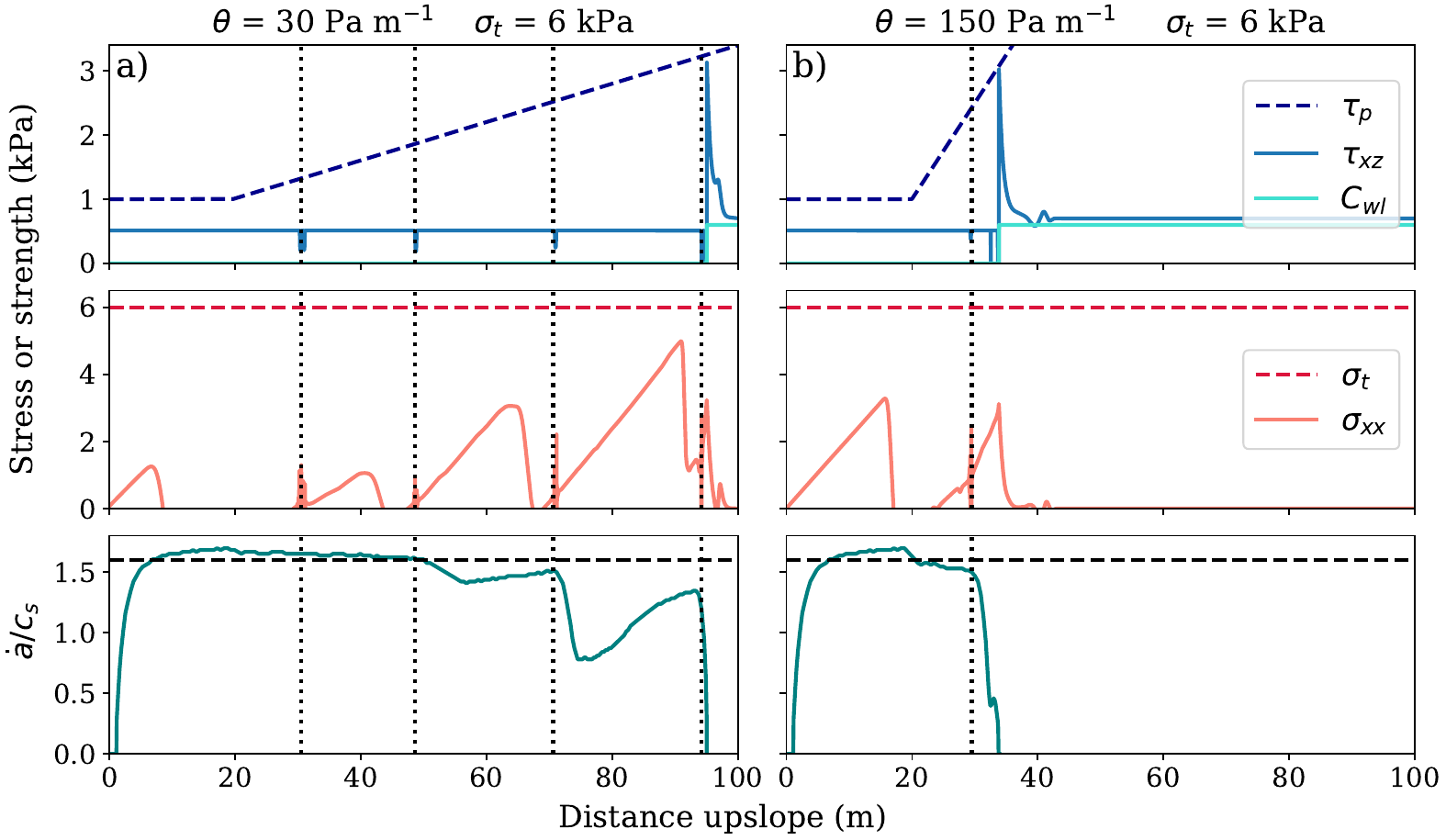}
\caption{Two PST simulations with elasto-plastic snow slabs. The top subplots show the shear stress $\tau_{xz}$ and strength $\tau_p$ as well as the weak layer cohesion $C_{wl}=\tau_p-\tau_r$. The middle subplots show the tensile stress $\sigma_{xx}$ and strength $\sigma_t$. The bottom subplots show the crack speed $\dot a$ normalised by the shear wave speed of the slab $c_s$. A horizontal dashed line is set at 1.6 $c_s$ for the crack speed. Vertical dotted lines represent locations where slab tensile fractures occurred along the PST. a) PST simulation with shear strength gradient $\theta$ = 30 Pa m$^{-1}$ with multiple tensile fracture (dotted lines) causing a crack arrest at 94 m. b) PST simulations with $\theta$ at 150 Pa m$^{-1}$ with only one tensile fracture causing a crack arrest at 34 m. In both simulation, the tensile strength $\sigma_t$ was set to 6000 Pa, and the softening coefficient $\delta$ to 0 (brittle).}
\label{fig:PSTtensile}
\end{figure}

The PST simulations presented in this section are performed with an elasto-plastic snow slab, allowing for slab tensile fractures to occur and potentially affect the propagation and cause the arrest of the crack. Simulations performed with a very low shear strength gradient, at 10 Pa m$^{-1}$, led to full propagation with slab tensile fractures that had no effect on the crack speed. Shear strength gradients of more than 20 Pa m$^{-1}$ were needed to obtain crack arrest within the PST length. In this cases, the crack arrest lengths were significantly lower than those performed with pure elastic slabs for the same weak layer shear strength gradient. Figure \ref{fig:PSTtensile} illustrates two PST simulations that resulted in different types of crack arrest. Supplementary videos are also provided to better grasp the spatio-temporal evolution of the stress states and crack dynamics. In Figure \ref{fig:PSTtensile}-a, the first type is achieved with a relatively low $\theta$ (30 Pa m$^{-1}$), and was characterized by four slab tensile fractures needed to stop crack propagation. Initially, the first slab fracture minimally affected the crack speed as the crack tip just passed 30 m (20 m = $L_{ss}$). However, subsequent slab fractures increasingly affected the crack speed until the last fracture completely stopped the propagation at around 94 m. Figure \ref{fig:PSTtensile}-b shows a PST simulation with a higher shear strength gradient $\theta$ of 150 Pa m$^{-1}$, yet with the same tensile strength $\sigma_t$ as in Figure \ref{fig:PSTtensile}-a. The higher value of $\theta$ caused the crack speed to decelerate significantly after the first slab fracture. In this case, only one slab tensile fracture was needed to stop crack propagation. Both the shear strength gradient $\theta$ and slab tensile fracture $\sigma_t$ contribute to the arrest of the crack. 

\begin{figure}[h!]
\includegraphics[width=\textwidth]{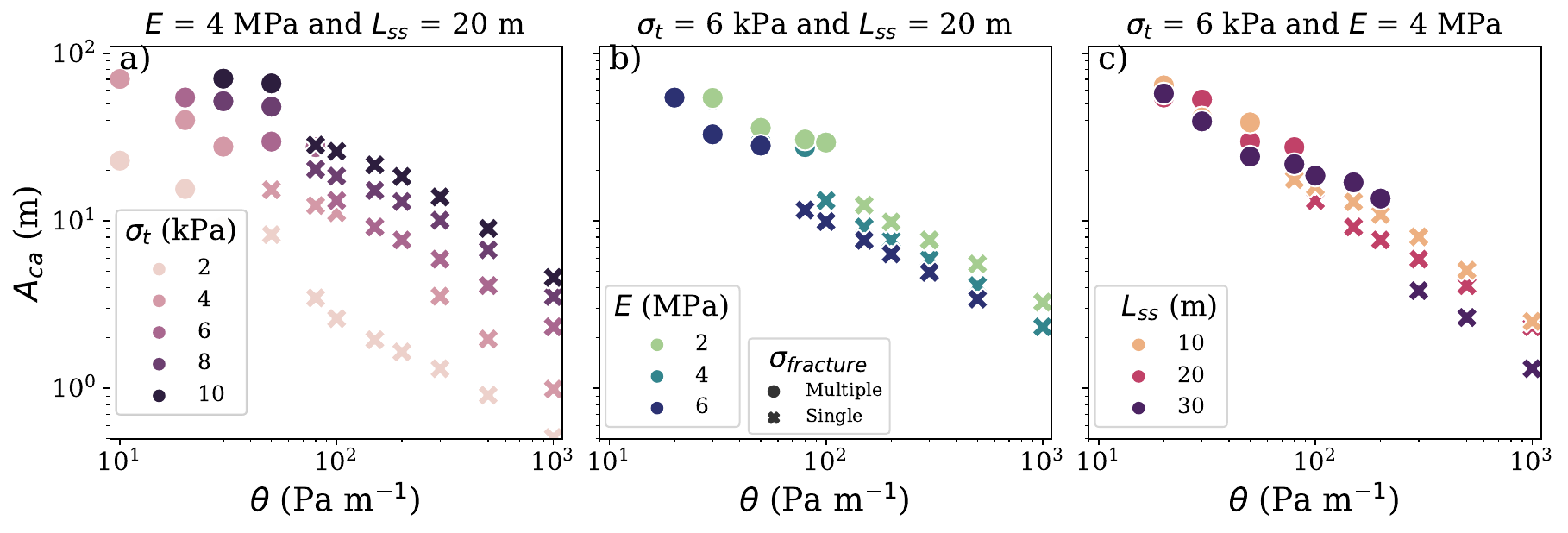}
\caption{Crack arrest length $A_{ca}$ obtained from PST simulations as a function of the shear strength gradient $\theta$, and for different a) tensile strengths $\sigma_t$ ($E$ = 4 MPa and $L_{ss}$ = 20 m), b) elastic modulus values $E$ ($\sigma_t$ = 6 kPa and $L_{ss}$ = 20 m), and c) steady-state lengths $L_{ss}$ ($\sigma_t$ = 6 kPa and $E$ = 4 MPa). These PST simulations were done with a softening coefficient $\delta$ of 1. The symbols represent cases in which there was only one slab tensile fracture (cross) needed to stop the crack and in which multiple slab fracture were needed (circles).}
\label{fig:PSTLca_theta}
\end{figure}

Figure \ref{fig:PSTLca_theta} presents a sensitivity analysis for four parameters in the case with an elastoplastic slab: shear strength gradient $\theta$, the tensile strength $\sigma_t$, the elastic modulus $E$ and the steady-state length $L_{ss}$. First, similar to the pure elastic case, we report a significant decrease of the crack arrest length $A_{ca}$ with increasing shear strength gradient, as expected. However, compared to the pure elastic slab case, values of $\theta$ about an order of magnitude lower were required with an elasto-plastic slab to obtain a similar crack arrest length. This result illustrates the complex interplay between slab tensile fractures and the crack dynamics in the weak layer on crack arrest propensity. 

In addition, the tensile strength $\sigma_t$ also had a significant influence on crack arrest. For a given $\theta$, a higher $\sigma_t$ led to slab tensile fractures occurring further upslope in the PST, resulting in longer $A_{ca}$  (Figure \ref{fig:PSTLca_theta}-a). The slab's elastic modulus $E$ also influences $A_{ca}$, where lower $E$ values correspond to larger $A_{ca}$ values for identical $\theta$ values.
The influence of the steady-state length $L_{ss}$ on $A_{ca}$ depends on the value of $\theta$. For large $\theta$ values, we report a decrease of $A_{ca}$ with increasing  $L_{ss}$ (Figure \ref{fig:PSTLca_theta}-c). However, for lower values of $\theta$, it appears that the effect of $L_{ss}$ becomes more moderate and tends to vanish with decreasing values of $\theta$.

Furthermore, this elasto-plastic analysis reveals two distinct types of crack arrest (Figure \ref{fig:PSTLca_theta}): crack with single or multiple slab tensile fractures. Multiple slab tensile fractures (circles in Figure \ref{fig:PSTLca_theta}), often called ''en-echelon'' fractures in the literature \citep{Gauthier2016OnAvalanches.},  occurred within the lower range of $\theta$. This regime of multiple fractures is also affected by several parameters such as $\sigma_t$, $E$ and $L_{ss}$ (Figure \ref{fig:PSTLca_theta}). For larger values of the shear strength gradient $\theta$, typically above 100 Pa/m, only a single slab tensile fracture was required to arrest the crack (crosses in Figure \ref{fig:PSTLca_theta}). Finally, note that simulations with $\theta$ below 20 Pa/m were removed from this analysis because the crack in the weak layer fully propagated throughout the entire PST length, without arrest, even with multiple slab tensile fractures.

\begin{figure}[h!]
\includegraphics[width=0.85\textwidth]{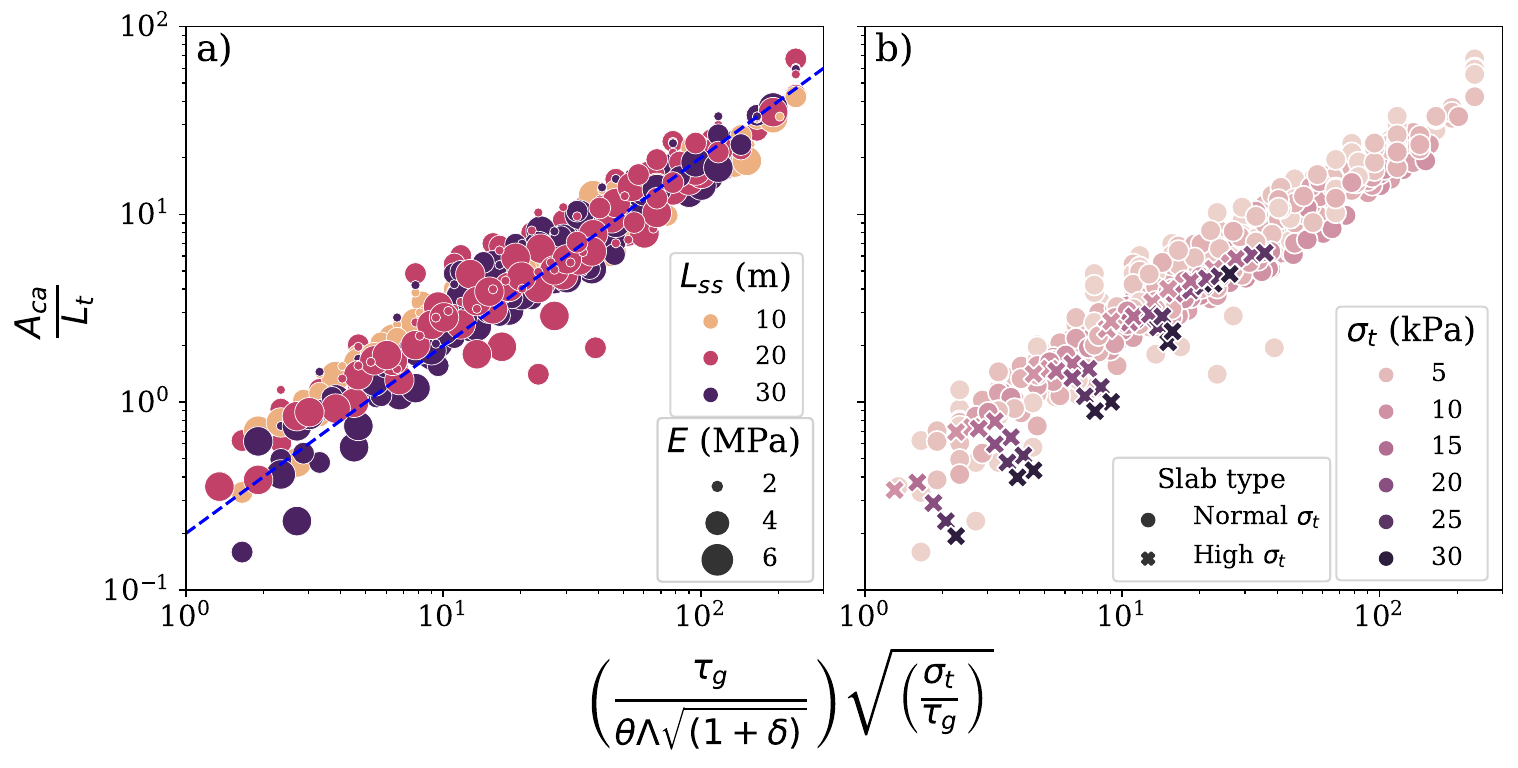}
\caption{Crack arrest length $A_{ca}$ (normalized with $L_{t} = \sigma_t / (\rho g \sin \psi (1- \frac{\tan \phi}{\tan \psi})) $) from all our PST elasto-plastic-simulations in relation to the product between two dimensionless numbers. The scaling takes into account all the parameters tested in this study, which are the shear strength gradient $\theta$, the tensile strength $\sigma_t$, the steady-state length $L_{ss}$, elastic modulus of the slab $E$,  and the softening coefficient $\delta$. a) PST elasto-plastic simulations with realistic values of $\sigma_t$ with respect to snow, with a 1:1 slope line in blue. b) PST elasto-plastic simulations with added simulations with very high $\sigma_t$ values, that demonstrated a transition regime from our elasto-plastic scaling to our pure elastic scaling.}
\label{fig:PSTLca_pi1}
\end{figure}

Numerous simulations were performed with various parameters to better understand the mechanisms driving crack arrest during snow slab avalanche release. Here, we simulated PSTs with different values of $L_{ss}$, $E$, $\sigma_t$, two different values of $\delta$ (perfectly brittle weak layer, i.e. $\delta$ = 0 and with a non-zero softening length i.e. $\delta$ = 1). On the basis of the large amount of simulation data generated, a second dimensional analysis was performed. The crack arrest length was normalized by the quasi-static tensile length $L_t = \sigma_t/(\rho g \sin \psi (1- \frac{\tan \phi}{\tan \psi}))$. This normalized crack arrest length $A_{ca}/L_{t}$ was plotted against the dimensionless number $\left(\frac{\tau_g}{\theta \Lambda\sqrt{1+\delta}}\right) \sqrt{\left(\frac{\sigma_t}{\tau_g}\right)}$. The left side of this number is the same as in the elastic scaling, and characterizes the ratio between the driving forces, namely the shear stress $\tau_g$ to a term characterizing the resisting forces, and related to $\theta$, $\Lambda$ and $\delta$. The right side of the term compares the slab tensile strength $\sigma_t$ to the slope parallel shear stress $\tau_g$. A low $\sigma_t/\tau_g$ ratio implies a low crack arrest length $A_{ca}$. We report in Figure \ref{fig:PSTLca_pi1}-a the result of this scaling analysis showing in log-scale a linear data collapse, thus suggesting a power-law relationship of the following form:
\begin{equation}
    \frac{A_{ca}}{L_{t}}\propto \left(\frac{\tau_g}{\theta \Lambda \sqrt{1+\delta}}\right) \sqrt{\left(\frac{\sigma_t}{\tau_g}\right)}
\end{equation}
Note that simulations with large values of tensile strength in which the arrest of the crack occurred without slab tensile fracture (arrest purely due to weak layer strength variations) were added to Figure \ref{fig:PSTLca_pi1}-b. These data points deviate significantly from the previously reported data collapse. Likewise, in Figure \ref{fig:PSTaca_purescale}-b, we introduced additional data points with slab fractures atop the scaling law established for an elastic slab. Once more, it appears that distinct mechanisms are in action, with varying factors influencing arrest in cases with and without slab fractures, as these additional data points fall outside the range of the data collapse out of the data collapse.

\subsection{Application to a real case with shear strength gradient $\theta$ measured \textit{in-situ}}
\begin{figure}[h!]
\includegraphics[width=0.9\textwidth, trim = 3.5cm 1cm 0.5cm 2cm]{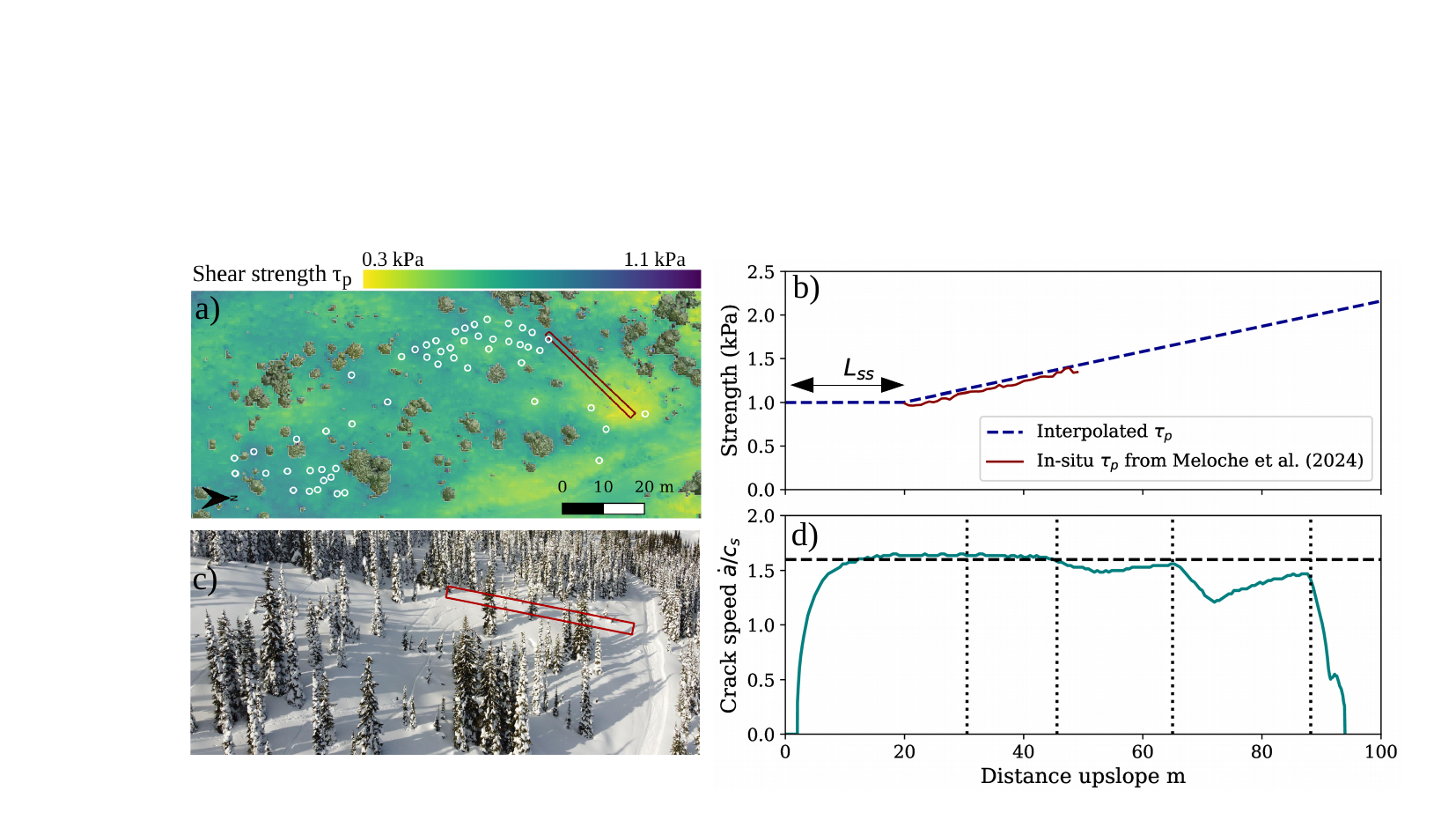}
\caption{PST simulation based on \textit{in-situ} measurement of shear strength $\tau_p$ and shear strength gradient $\theta$ according to measurements made by \cite{Meloche2024SnowModelling} at a forested site Jim Bay Corner. The other parameters were the same as for the elasto-plastic simulations described in Table \ref{table.param}, with $L_{ss}$ = 20 m, $E$ = 4 MPa, $\sigma_t$ = 6 kPa and $\delta$ = 1. a) Shear strength map of \cite{Meloche2024SnowModelling} at the Jim Bay corner, with the sample location in the red rectangle. b) \textit{In-situ} shear strength $\tau_p$ fitted at 1 kPa with the interpolated gradient $\theta$, note that the shear strength values were fitted to start at 1 kPa in order to match our simulation parameters. c) Aerial photography of the site from UAV imagery. d) Crack speed $\dot{a}$ normalized by $c_s$ from the simulation with slab tensile fractures (vertical dotted lines).}
\label{fig:JBCinsitu}
\end{figure}

\begin{figure}
\includegraphics[width=0.95\textwidth, trim = 3.5cm 1cm 0.5cm 2cm]{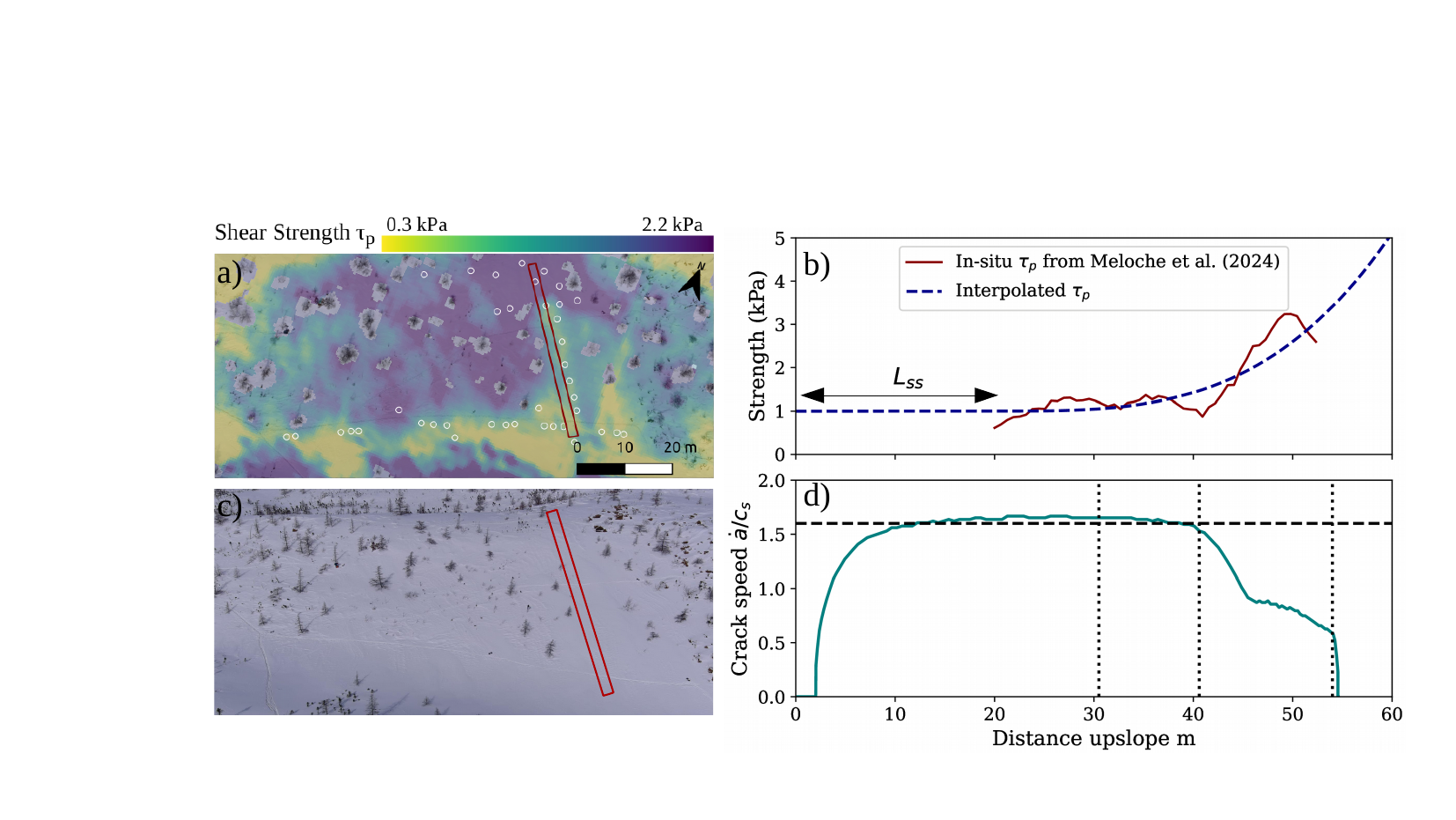}
\caption{PST simulation based on \textit{in-situ} measurement of shear strength $\tau_p$ and shear strength gradient $\theta$ according to measurements made by \cite{Meloche2024SnowModelling} at an alpine site called Arête de Roc. The other parameters were the same as for the elasto-plastic simulations described in Table \ref{table.param}, with $L_{ss}$ = 20 m, $E$ = 4 MPa, $\sigma_t$ = 6 kPa and $\delta$ = 1. a) Shear strength map of \cite{Meloche2024SnowModelling} at the Arête de Roc, with the sample location in the red rectangle. b) \textit{In-situ} shear strength $\tau_p$ fitted at 1000 Pa with the interpolated gradient $\theta$, note that the shear strength values were fitted to start at 1000 Pa in order to match our simulation parameters. c) Aerial photography of the site from UAV imagery. d) Crack speed $\dot{a}$ normalized over $c_s$ from the simulation with slab tensile fractures (vertical dotted lines).}
\label{fig:ARinsitu}
\end{figure}

The values of $\theta$ used in the previous elasto-plastic simulations were selected in order to obtain crack arrest within our 100 m PST length. In this section, instead, model parameters are input on the basis of field measurements. \cite{Meloche2024SnowModelling} performed \textit{in-situ} measurements of snow mechanical properties in order to derive snow strength and stability maps at the slope scale. We sampled their weak layer shear strength maps in the slope parallel direction from a weaker area to the top of the slope to obtain a shear strength gradient. Figures \ref{fig:JBCinsitu}-a and \ref{fig:ARinsitu}-a show the sample location of the gradient (red rectangles) in two different study sites. We then interpolated the \textit{in-situ} gradient in order to get a longer distance that will match our 100 m PST length. We applied these interpolated gradient after a steady-state length of 20 m to obtain the same elasto-plastic PST simulation set-up but with the \textit{in-situ} gradient (Figures \ref{fig:JBCinsitu}-c and \ref{fig:ARinsitu}-c). The first \textit{in situ} $\theta$ value was obtained at a forested site called Jim Bay corner JBC, and exhibited a linear gradient of 15 Pa m$^{-1}$. The crack speed from the simulation with the JBC gradient is shown in Figure \ref{fig:JBCinsitu}-d, demonstrating that four slab fractures (vertical dotted lines) were needed to stop the crack propagation. This \textit{in-situ} gradient from JBC is rather small and is located in the lower range of the $\theta$ values tested in this study.

The second \textit{in-situ} shear strength was sampled in a wind-affected alpine site from \cite{Meloche2024SnowModelling}. The shear strength map had more important variations as well as larger maximum shear strength values compared to the first site. This second site exhibited a nonlinear gradient, and a power-law was fitted to these values in order to apply the nonlinear gradient to match our longer PST (Figure \ref{fig:ARinsitu}-b). This simulation needed three slab fractures in order to stop the propagation. The first slab fracture did not affect the propagation in the weak layer. However, the second fracture, which occurred at the same location where the gradient is starting to increase rapidly, significantly slowed down the crack speed below $c_s$ at around 40 m. A third slab fracture was needed to completely stop the crack propagation at 55 m. This nonlinear gradient observed at the alpine site was much larger than that obtained in the previous forested site, thus leading to a shorter crack arrest distance of around 55 m.

\subsection{Slope scale analysis on three-dimensional terrain}
\begin{figure}
\includegraphics[width=0.8\textwidth, trim = 0cm 0cm 0cm 0cm]{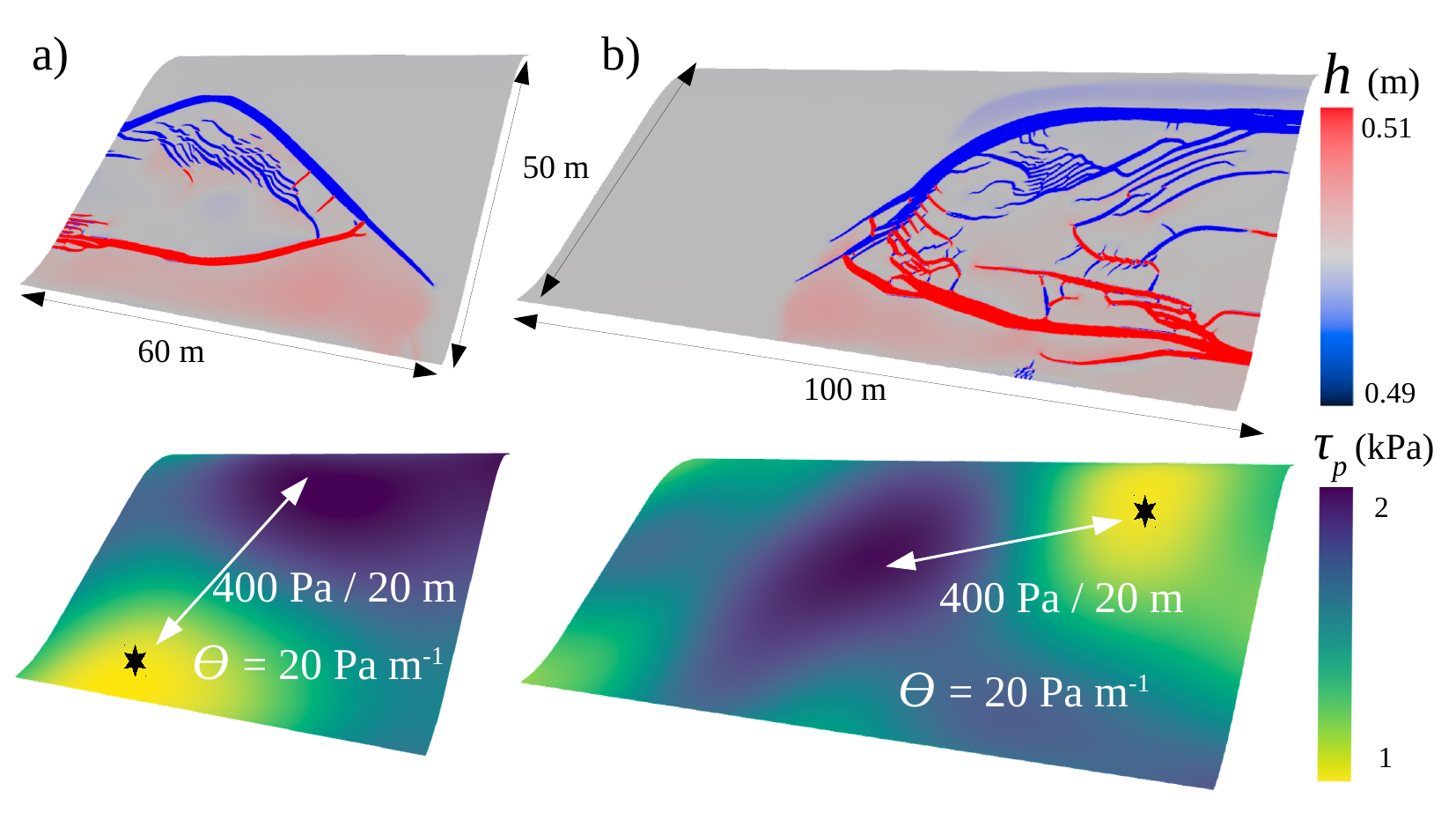}
\caption{Slope scale crack propagation simulation with weak layer heterogeneity mostly oriented in the a) slope-parallel direction and in the b) cross-slope direction. The particle height $h$ represent the tensile fractures ($h<0.49$) and the compressive fractures ($h>0.51$) within the slab. The shear strength gradient was generated with Gaussian Random Field GRF, corresponding to a $\theta$ of 20 Pa m$^{-1}$ in slope parallel direction and cross-slope direction, a slab elastic modulus $E$ of 4 MPa, and a tensile strength of 4 kPa. We initiated a crack in the weak zone, denoted by a black star, at the weaker area with the same procedure as in the PST.}
\label{fig:fullscale}
\end{figure}

The last result section is to explore the slope scale crack propagation with a weak layer heterogeneity in both slope parallel and the cross-slope direction. Figure \ref{fig:fullscale}-a demonstrates the applicability of the model on a fictional slope of 50 m by 60 m wide. Using a Gaussian Random Field approach (GRF), a 2D natural variability was generated, showing a shear strength gradient of 20 Pa m$^{-1}$, with an increase of around 400 Pa over 20 m up-slope (Figure \ref{fig:fullscale}-a). We initiated a crack at the bottom of the slope in the weak zone (black star) which propagated up-slope and stopped near the upper and stronger part of the slope. The DA-MPM column height was used to visualize slab tensile fractures in blue ($h < 0.49$), and the compressive fractures within the slab in red ($h > 0.51$). The most important slab tensile failure in blue represents the so-called avalanche crown where the crack propagation stopped. In contrast, the most important area in red (compressive failure) represents the so-called stauchwall of the avalanche. The crack propagation also stopped in the cross-slope direction at the right corner of the slope. Figure \ref{fig:fullscale}-a shows that crack propagation stopped between 40-50 m, approximately the same values of $A_{ca}$ that were obtained in our PST experiment for the same value of $\theta$ (Figure \ref{fig:PSTLca_theta}). However, the crown is also starting at the beginning of the artificial convex roll where the slope angle is starting to reduce. The arrest of the crack might be caused by a combination of the increase in shear strength but also by the decrease in slope angle. Furthermore, Figure \ref{fig:fullscale}-b is showing a crack propagation in the cross-slope direction with a weak layer heterogeneity mostly oriented in the cross-slope direction. The initiation was at the top right corner of the slope (black star), and it propagated in the cross and down slope direction until the middle of the slope where the shear strength locally increases. The arrest in the cross-slope direction can only be related to the heterogeneity of the weak layer because the slope angle is constant in that part of the slope. However, it is known that the crack propagation speed in the cross-slope direction is lower than the downslope propagation speed, and is approximately around the shear wave speed of the slab $c_s$ \citep{Guillet2023ARelease,Trottet2022TransitionAvalanches, Simenhois2023UsingSpeeds,Gaume2019InvestigatingMethod}. This slower speed in crack propagation is easier to stop, as shown in the previous section. In the up-slope direction, terrain variations (reduction in slope angle) induce the arrest of the crack and the slab tensile failure.

\section{Discussion}
In this study, we investigated the mechanical drivers of crack arrest during dynamic crack propagation in snow slab avalanche release. For this purpose, we used a depth-averaged Material Point Method (DA-MPM, \cite{Guillet2023ARelease}) which relies on a strain-softening shear failure model for the weak layer, and an elasto-plastic model for the overlaying slab. Two types of crack arrest were investigated: i) crack arrest induced by heterogeneity of weak layer mechanical properties (here the strength) and ii) crack arrest induced by slab fracture. 

Initial findings with a pure-elastic slab regarding the effect of softening and shear strength gradient were rather intuitive, but had not been quantified yet. An increasing weak layer shear strength gradient implied a decreasing crack arrest length, as expected. In addition, we showed that increasing weak layer fracture energy dissipated during the softening phase, resulted in reduced propagation distances. We also report a strong dynamic effect on the slab tensile stress state with higher crack propagation speeds leading to lower slab tensile stress compared to quasi-static predictions. This stress reduction, observed in numerical simulations has been verified on the basis of analytical developments (see Appendix A). This result suggests that fast cracks in the weak layer are likely to produce larger release areas. In the context of the anticrack to supershear transition reported by \cite{Trottet2022TransitionAvalanches} and \cite{Bobillier2024NumericalExperiments,Bobillier2024SupershearMeasurements}, this also implies that slab fractures in the supershear regime are likely to occur at larger distances from the initiation point compared to the anticrack propagation regime which involve much lower crack speeds (below $c_s$).
\cite{Gaume2015ModelingMethod} made similar observations with smaller PST experiment of 2 m long in DEM. They observed a lower tensile stress in dynamic crack propagation compared to the quasi-static tensile stress. This resulted in a longer propagation distance compared to predictions from beam theory. Furthermore, although they did not study the interplay between crack dynamics and slab fracture in details, \cite{Gaume2015InfluenceArea} reported a reduction in slab tensile fracture propensity with increasing slab density (and thus stiffness). They attributed this reduction to smoothing effects which could however be exacerbated by the dynamic effects revealed here. 

In a second step, PST simulations with an elasto-plastic slab demonstrated that crack arrest could occur for much lower shear strength gradients compared to pure elastic slabs because of the occurrence of slab fracture. More precisely, slab fracture significantly reduced the crack arrest length for the same the shear strength gradient. Although the positive shear strength gradient induced a reduction of the crack speed, slab fracture had the most significant effect and appeared to be the main driver of crack arrest (Figure \ref{fig:PSTtensile}). In numerous of the presented simulations, multiple slab fractures were needed to stop the crack and the number of slab fractures decreased with increasing shear strength gradient. These multiple fractures are reminiscent of the ''en-echelon'' slab fracture mechanism reported in \cite{Gauthier2016OnAvalanches.} and further discussed in \cite{Gaume2019InvestigatingMethod,Trottet2022TransitionAvalanches}. It is important to note that slab fractures alone are insufficient to arrest the crack in this supershear regime, as reported by \cite{Trottet2022TransitionAvalanches}. It is really the interplay between slab fracture and spatial variability that ultimately contribute to crack arrest. In our study, heterogeneity is represented by a weak layer strength increase. In principle, variability of topography and/or other mechanical properties could induce a similar effect and should be studied in the future. 

%tensile fracture
The stopping of the crack propagation by tensile fracture was previously observed by many in smaller PST experiment \citep{Birkeland2014ProceedingsAlberta, Gaume2015ModelingMethod, vanHerwijnen2005FracturesRelease, Bergfeld2023TemporalPropagation}. However, the crack arrest conditions explored in our study are not comparable to the ones reported in smaller PST experiments by the latter authors. The tensile fracture reported in their work originated mostly from the top of the slab caused by the bending of the slab due to collapse weak layer. Because our model relies on a pure shear weak layer failure model, following ideas of \cite{Guillet2023ARelease} and results of \cite{Trottet2022TransitionAvalanches}, our model, by nature, cannot reproduce such observations.
However, \cite{Gaume2019InvestigatingMethod} observed in 3D MPM slab avalanche simulations that some slab fractures originated from the bottom of the slab. \cite{Trottet2022TransitionAvalanches} corroborated these observations and added that this slab tensile fracture were related to the shear band propagation in the supershear propagation regime. Crack arrest resulting from slab fractures branching from the bottom of the slab was also reported in \cite{Bair2016TheResearch} and \cite{Mcclung2021ApplicationRelease} based on avalanche crown observations. Hence, our study focuses on such slab fracture originating from weak layer shear band propagation and influencing wide-spread propagation and not on slab fractures reported at small-scale related to the anticrack mechanism. 

%application

The applicability of the proposed model was tested in three ways. First, scaling laws for the crack arrest length were proposed and could be implemented in operational snow cover and forecasting models as avalanche size indicators \citep{Durand1999AForecasting,Lehning1999SNOWPACKStations,Reuter2018DescribingSupport}. Second, it was applied to PST cases in which spatial variability of weak layer shear strength was obtained directly from \textit{in-situ} measurements \citep{Meloche2024SnowModelling}. Third, simulations over 3D topography with realistic weak layer strength spatial variability. While most of the present study focused on up-slope and mode II crack propagation, these simulations showed significantly lower crack propagation speeds in the cross-slope direction and thus in mode III. This corroborates recent findings \citep{Guillet2023ARelease,Trottet2022TransitionAvalanches, Simenhois2023UsingSpeeds} as well as theoretical bounds \citep{Broberg1989TheVelocities}. Although these applications bring us a step closer towards estimating avalanche release sizes, further research is needed to investigate the complex three-dimensional interplay between mode II (up/down-slope), mode III (cross-slope) shear band propagation, slab fracture and crack arrest in order to fully understand the mechanical and geometrical drivers of the shape and size of avalanche release zones.

Regarding limitations, we recall that the nature of the numerical (depth-averaged) and mechanical (weak layer mode II failure) models employed in this study only allows us to simulate the supershear crack propagation regime \citep{Trottet2022TransitionAvalanches} and  prevents us from modeling weak layer volumetric collapse and associated slab bending. As a consequence, our model is well suited to evaluate the release sizes of rather large avalanches, and may overestimate the size of small soft slabs with lower densities. The validity of this DA-MPM model for soft slab avalanche modeling could be further assessed by comparing results with 3D MPM which can simulate both anticrack and supershear crack propagation regimes. However, in view of risk management, we believe that this model can provide conservative estimates that can be used as input of avalanche dynamics models for hazard mapping purposes. In fact, in principle, the present model could be extended to simulate, in a unified way, both the release and the flow of the avalanche, in a similar manner as in \cite{Zhang2021DepthEvolution} for submarine landslides. The model already captures slab fragmentation after the release of the slab but additional work is required to modify the extent of the material to simulate after the release (deletion of particles outside the release zone), the rheological model which should incorporate rate-dependent behavior \citep{Gaume2020MicroscopicMaterials, Blatny2023ModelingApproach} as well as entertainment and centrifugal laws.

Finally, although the present application focused on crack arrest mechanisms in snow slab avalanches, similar analyses have been proposed to evaluate failure extension in landslides \citep{Zhang2021UpslopeRetrogression} and earthquakes \citep{Barras2023HowRupture} using closed-form criteria and analytical solutions. With process-specific modifications, we believe that the model presented here could be broadly applicable to other geophysical processes as well, providing a robust framework for understanding and predicting failure dynamics across a range of natural phenomena.

\section{Conclusion}

This study provides a comprehensive exploration of dynamic crack propagation on inclined slopes, particularly focusing on factors affecting crack arrest in the recently discovered supershear propagation regime. By initially examining pure-elastic slabs and then incorporating elasto-plastic behavior, critical factors such as the softening coefficient, shear strength gradient, crack propagation speed and steady-state distance were identified as influential in determining crack behavior and arrest distances. Importantly, the study revealed that crack dynamics and arrest mechanisms differ significantly from quasi-static predictions, with implications for understanding slab fracture propensity and avalanche release sizes. By conducting sensitivity analysis, proposing scaling laws and performing field data-informed simulations over 2D and 3D topographies, the research offers valuable insights for potential avalanche release size parameterizations based on terrain and snow properties. While highlighting practical implications for avalanche forecasting and risk assessment, the study also points to the need for further investigation, particularly in understanding crack arrest drivers in different fracture propagation modes and in complex 3D scenarios. Ultimately, this study advances our understanding of crack propagation dynamics and arrest in snow slab avalanches and lays the groundwork for future research in this field, contributing to avalanche risk management. For instance, the findings could inform the development of targeted mitigation strategies in avalanche release zones, aimed at preventing widespread crack propagation and reducing the avalanche size and its impact.

%% The following commands are for the statements about the availability of data sets and/or software code corresponding to the manuscript.
%% It is strongly recommended to make use of these sections in case data sets and/or software code have been part of your research the article is based on.

%%\codeavailability{TEXT} %% use this section when having only software code available

%%\dataavailability{TEXT} %% use this section when having only data sets available

\codedataavailability{The code and the data are available upon request.} %% use this section when having data sets and software code available

%%\sampleavailability{TEXT} %% use this section when having geoscientific samples available

%%\videosupplement{TEXT} %% use this section when having video supplements available

%% Please add \clearpage between each table and/or figure. Further guidelines on figures and tables can be found below.

\authorcontribution{FM, GB and JG conceptualized the research. FM led the research, ran the simulations and drafted the manuscript, with a close supervision and help of GB. LG wrote the DA-MPM code, provided useful insights on the numerical simulations, and derived the analytical solution for the understress regime. JG provides significant suggestions and ideas throughout the whole research process, review the manuscript and funded a part of the research. FG and AL funded the research and reviewed the manuscript.}

\competinginterests{} %% this section is mandatory even if you declare that no competing interests are present

%%\disclaimer{TEXT} %% optional section

\begin{acknowledgements}
This project was funded by the Quebec Research Funds - Nature and Technologies (FRQNT) and the Swiss National Science Foundation. We would like to thank all members of the chair of Alpine Mass movements ALMO, and also Guillaume Meyrat at SLF, who shared ideas during team meetings and coffee breaks.
\end{acknowledgements}

\appendix
\section{Analytical solution for the "controlled" regime with purely elastic slab}

% I try not to be very specific, maybe a part can go in the discussion
Let's consider a classic one dimensional model. We recall that we qualify the regime as "controlled" if the shear strength of the WL tends toward infinity (which means that there is no critical length) and the speed of the crack is manually imposed and constant. The failure only occurs by setting the cohesion to 0 (damaged weak layer) and the shear stress to the residual friction stress $\tau_r$. Even in this case, the time derivative of the displacement is not null and we need to consider the dynamic equation instead of the quasi-static ones, more commonly used. The displacement $u(x, t)$ must satisfy
\begin{equation}
    -\frac{1}{c_p^2} \frac{\partial^2 u}{\partial t^2} + \frac{\partial^2 u}{\partial x^2} = \frac{\tau_g}{E'h} + \frac{\tau(u)}{E'h}.
    \label{dynamic}
\end{equation}
where $c_p = \sqrt{\frac{E'}{\rho}}$ is the longitudinal wave speed in 1 dimension and $E'$ is the apparent Young's modulus. The function $\tau$ is defined in the "controlled" regime by 
\begin{align}
    \tau(u(x, t)) &= \tau_r \text{ if } x \leq \dot{a} t \text{ ( Zone 1 : No Cohesion)} \\
    \tau(u(x, t)) &= K_{_{WL}} u(x, t) \text{ if } x > \dot{a} t \text{ (Zone 2 : Cohesion)}.
\end{align}
We can then rewrite Eq.~(\ref{dynamic}) as
\begin{align}
    \label{Dyn_1}
    \text{if } x \leq \dot{a} t, \quad &-\frac{1}{c_p^2} \frac{\partial^2 u}{\partial t^2} + \frac{\partial^2 u}{\partial x^2} = \frac{k_f}{E'} \\
    \label{Dyn_2}
    \text{if } x > \dot{a} t, \quad &-\frac{1}{c_p^2} \frac{\partial^2 u}{\partial t^2} + \frac{\partial^2 u}{\partial x^2} = \frac{1}{\Lambda^2}u(x, t) + \frac{\tau_g}{E'h}
\end{align}
with $k_f = \frac{\tau_g - \tau_r}{h}$ the quasi-static stress slope and $\Lambda = \sqrt{\frac{E'h}{K_{_{WL}}}}$ the caracteristic elastic length. The differential system given by Eqs.~(\ref{Dyn_1}) and (\ref{Dyn_2}) is highly nonlinear and by itself difficult to solve. However with two minor assumptions we can compute an analytical solution for the stress in Zone 1.
\begin{itemize}
    \item In Zone 1, the stress has a constant slope in space : the second space derivative $\frac{\partial^2 u}{\partial x^2}$ and depends only on $\dot{a}$ and $k_f$. 
    \item The displacement at the frontier is proportional to the time : $u(\dot{a} t, t) \propto t$
\end{itemize}

Note that these two hypotheses do not rely on purely mathematical development but are very reasonable regarding the physical model and the output of our simulations. We then denote by $k$ the stress slope in zone 1, $k_x = E' \frac{\partial^2 u}{\partial x^2}$. By integrating two times and injecting in Eq.~(\ref{Dyn_1}) we have that 
\begin{equation}
    u(x, t) = \frac{1}{2}k_x x^2 + \frac{1}{2} c_p^2 (k_x - k_f)t^2 + \alpha_1 t + \alpha_2
\end{equation}
where $\alpha_1$ and $\alpha_2$ are two integration constant. The second hypothesis gives that
\begin{equation}
    \frac{1}{2}k_x (\dot{a}t)^2 + \frac{1}{2} c_p^2 (k_x - k_f)t^2 + \alpha_1 t + \alpha_2 \propto t
\end{equation}
which implies that the term before $t^2$ must be equals to zero. It then reduces to 
\begin{equation}
    k_x \dot{a}^2 + \frac{1}{2} c_p^2 (k_x - k_f) \iff k_x = \frac{c_p^2k_f}{\dot{a}^2 + c_p^2}
\end{equation}

Note that when $\dot{a} = c_p$, the slope is just half as the slope in quasi-static regime with $k_x = \frac{k_f}{2}$. Even if the 'controlled' regime has no real physical value when $\dot{a}$ is comparable to $c_p$, it can gives a strong insight of what must be the value of convergence of the stress slope in self propagating regime.

%% REFERENCES

%% The reference list is compiled as follows:

%\bibliographystyle{copernicus}
%\bibliography{references.bib}

\end{document}